\documentclass[aps,reprint,groupedaddress]{revtex4-1}

\usepackage{amsmath}
\usepackage{amssymb}
\usepackage{mathrsfs}

\usepackage{txfonts}
\usepackage{graphicx}

\DeclareMathOperator{\Tr}{\mathop{Tr}}

\begin{document}

\title{Frustrations in the Ising chain with the third-neighbor interactions}

\author{A. V. Zarubin}
\email{Alexander.Zarubin@imp.uran.ru}

\author{F. A. Kassan-Ogly}
\author{A. I. Proshkin}

\affiliation{M. N. Mikheev Institute of Metal Physics of Ural Branch of Russian Academy of Sciences, S. Kovalevskoy Street 18, 620108 Ekaterinburg, Russia}

\date{\today}

\begin{abstract}
We study the frustration properties of the Ising model on a one-dimensional
monoatomic equidistant lattice, taking into account the exchange interactions
of atomic spins at the sites of the nearest, next-nearest, and third
neighbors. The exact analytical expressions for the thermodynamic
functions of the system are obtained using the Kramers--Wannier transfer
matrix technique. Criteria for the emergence of magnetic frustrations
in the presence of competition between the energies of exchange interactions
are formulated. The points and intervals of the existence of frustrations,
which depend on the values and signs of the exchange interactions,
are found. The features of the entropy and heat capacity of this model
in the frustration regime and its vicinity are investigated. Non-zero
entropy values of the ground state of a frustrated system, as well
as a two-peak temperature structure of the heat capacity in the vicinity
of the frustration point, are found.
\end{abstract}

\maketitle

\section{Introduction}

Spin systems with magnetic frustrations are a rapidly developing field
of research in recent years, which covers a wide range of objects
with special magnetic states, such as spin liquid, spin ice, as well
as various incommensurate, helicoidal, chiral, and other exotic structures~\citep{Kassan-Ogly:2010:,Normand:2009,Balents:2010,Lacroix:2011,Sadoc:1999,Kudasov:2012:,Diep:2013,Vasiliev:2018:}.

The phenomenon of magnetic frustration was discovered in the mid-seventies
of the twentieth century in magnetic materials exhibiting unusual
properties, which was explained by the strong degeneration of the
ground state of the system and the impossibility of magnetic ordering
even at zero temperature. Such magnetics by Gerard Toulouse in 1977
were called \emph{frustrated}~\citep{Toulouse:1977:1,Toulouse:1977:2}.

The experimental material on frustrated magnetic systems in real crystals
and noncrystalline materials is very rich and is replete with new
phenomena and unusual properties. However, the proper interpretation
and theoretical explanation of a multitude of experimental facts and
new effects is currently absent, and a number of properties of the
frustrated systems are not yet sufficiently understood.

In the present paper, we study the frustration properties of the one-dimensional
Ising model on a monatomic equidistant lattice taking into account
the exchange interactions of atomic spins on the sites of the nearest,
next-nearest, and third neighbors. This model makes it possible to
obtain an exact solution in the thermodynamic limit, and qualitatively
consider the desired characteristics, including explaining the properties
of magnetic materials caused by frustrations, which are not available
for description within perturbation theory~\citep{Baxter:1982}.

Of course, the Ising model has long been widely used in the theory
of magnetism and has a set of well-known solutions \citep{Ising:1925,Brush:1967,Niss:2005,Newell:1953},
but no systematic description of its frustration properties has been
carried out.

Thus, the computation of the thermodynamic characteristics of the
Ising model allows one to find the essential information about the
frustration of the system~\citep{Kassan-Ogly:2001,Kassan-Ogly:1989:,Zarubin:2016}
and relate it to the experimental observables.

\section{Thermodynamic functions of the Ising chain}

We will consider the one-dimensional classical Ising model taking
into account the exchange interactions between atomic spins at the
sites of the first (nearest), second (next-nearest), and third neighbors,
which is given by the Hamiltonian
\begin{equation}
\mathscr{H}=-\sum_{p=1}^{b}\sum_{n=1}^{N-p}J_{p}\sigma_{n}\sigma_{n+p},\label{eq:H:0}
\end{equation}
where $b$ is the number of exchange interactions of the chain spins
in the model (in this case $b=3$), $J_{1}$ is the parameter of exchange
interaction between the spins at the nearest neighbor sites in the
linear lattice, $J_{2}$ is the parameter of exchange interaction
between the spins at the next-nearest lattice sites, $J_{3}$ is the
parameter of exchange interactions between the spins at the third
neighbors, the symbol $\sigma_{n}$ denotes the $z$ projection of
the atom spin operator located at the $n$-site and is equal to $\sigma=\pm1$,
and $N$ is the number of the chain sites.

In the Kramers--Wannier tansfer matrix method \citep{Kramers:1941:1,Baxter:1982}
used with the Born--von K\'{a}rm\'{a}n cyclic boundary conditions are imposed
\[
\sigma_{N+1}=\sigma_{1},
\]
the partition function is
\begin{equation}
Z=\Tr\mathbf{V}^{N},\label{eq:PF}
\end{equation}
where $\mathbf{V}$ is the transfer matrix the elements of which are
independent of the site index \citep{Baxter:1982} and are specified
by the rule
\begin{multline}
V_{\sigma'''\sigma''''\sigma'''''}^{\sigma\sigma'\sigma''}=\langle\sigma\sigma'\sigma''|e^{K_{1}\sigma\sigma'+K_{2}\sigma\sigma''+K_{3}\sigma\sigma'''}|\sigma'''\sigma''''\sigma'''''\rangle=\\
=e^{K_{1}\sigma\sigma'+K_{2}\sigma\sigma''+K_{3}\sigma\sigma'''}\delta_{\sigma'\sigma'''}\delta_{\sigma''\sigma''''}\label{eq:TM:V}
\end{multline}
through dimensionless quantities
\[
K_{1,2,3}=\beta J_{1,2,3},\quad\beta=\frac{1}{k_{\text{B}}T},
\]
and $\delta_{\sigma'\sigma''}$ is the Kronecker symbol,

Note that in further transformations, the Boltzmann constant $k_{\text{B}}$
will be put equal to unity, and the quantities $T$, $J_{2}$, and
$J_{3}$ will be measured in the units of $|J_{1}|$, as is commonly
accepted in the theory of low-dimensional systems. 

The dimension of the square transfer matrix of a one-dimen\-sional spin
model is determined by the expression
\begin{equation}
d=c^{b},\label{eq:TM:dim}
\end{equation}
where $c$ is the number of states at a site ($c=2$ in the classical
Ising model), and $b$ is the number of exchange interactions of spins
of the chain in the problem ($b=3$). Therefore, in the considered
problem, the dimension of the transfer matrix is equal to
\[
d=2^{3}.
\]

The construction of the transfer matrix was carried out according
to the scheme proposed in \citep{Oguchi:1965}, and described in detail
in \citep{Zarubin:2019:}. We obtain that the transfer matrix has
the following form
\begin{widetext}
\begin{equation}
\mathbf{V}=\left(\begin{array}{cccccccc}
e^{K_{1}+K_{2}+K_{3}} & e^{K_{1}+K_{2}-K_{3}} & 0 & 0 & 0 & 0 & 0 & 0\\
0 & 0 & e^{K_{1}-K_{2}+K_{3}} & e^{K_{1}-K_{2}-K_{3}} & 0 & 0 & 0 & 0\\
0 & 0 & 0 & 0 & e^{-K_{1}+K_{2}+K_{3}} & e^{-K_{1}+K_{2}-K_{3}} & 0 & 0\\
0 & 0 & 0 & 0 & 0 & 0 & e^{-K_{1}-K_{2}+K_{3}} & e^{-K_{1}-K_{2}-K_{3}}\\
e^{-K_{1}-K_{2}-K_{3}} & e^{-K_{1}-K_{2}+K_{3}} & 0 & 0 & 0 & 0 & 0 & 0\\
0 & 0 & e^{-K_{1}+K_{2}-K_{3}} & e^{-K_{1}+K_{2}+K_{3}} & 0 & 0 & 0 & 0\\
0 & 0 & 0 & 0 & e^{K_{1}-K_{2}-K_{3}} & e^{K_{1}-K_{2}+K_{3}} & 0 & 0\\
0 & 0 & 0 & 0 & 0 & 0 & e^{K_{1}+K_{2}-K_{3}} & e^{K_{1}+K_{2}+K_{3}}
\end{array}\right).\label{eq:N3:TM}
\end{equation}
\end{widetext}

The resulting matrix (\ref{eq:N3:TM}) can be reduced to the block
form, and the characteristic equation of which is defined as
\begin{multline}
(\lambda^{4}+a_{3}\lambda^{3}+a_{2}\lambda^{2}+a_{1}\lambda+a_{0})(\lambda^{4}+b_{3}\lambda^{3}+b_{2}\lambda^{2}+b_{1}\lambda+b_{0})=0,\label{eq:N3:CP1}
\end{multline}
where the coefficients are
\[
a_{3}=-2e^{K_{2}}\cosh(K_{1}+K_{3}),\quad b_{3}=-2e^{K_{2}}\sinh(K_{1}+K_{3}),
\]
\[
a_{2}=-b_{2}=2\sinh(2K_{2}),
\]
\[
a_{1}=4e^{-K_{2}}\sinh(2K_{3})\sinh(K_{1}-K_{3}),
\]
\[
b_{1}=-4e^{-K_{2}}\sinh(2K_{3})\cosh(K_{1}-K_{3}),
\]
\[
a_{0}=b_{0}=4\sinh^{2}(2K_{3}).
\]

The principal (single largest real) eigenvalue of the matrix (\ref{eq:N3:TM})
determined from the equation (\ref{eq:N3:CP1}), is expressed in radicals
and has the following form
\begin{equation}
\lambda_{1}=-\frac{a_{3}}{4}-\Psi+\frac{1}{2}\sqrt{-4\Psi^{2}-2p+\frac{q}{S}},\label{eq:L1}
\end{equation}
\[
p=a_{2}-\frac{3}{8}a_{3}^{2},\quad q=a_{1}-\frac{a_{2}a_{3}}{2}+\frac{a_{3}^{3}}{8},
\]
\[
\Psi=\frac{1}{2}\sqrt{-\frac{2}{3}p+\frac{1}{3}\left(\Theta+\frac{\Delta_{0}}{\Theta}\right)},
\]
\[
\Theta=\sqrt[3]{\frac{\Delta_{1}+\sqrt{\Delta_{1}^{2}-4\Delta_{0}^{3}}}{2}},
\]
\[
\Delta_{0}=12a_{0}-3a_{1}a_{3}+a_{2}^{2},
\]
\[
\Delta_{1}=-72a_{0}a_{2}+27a_{0}a_{3}^{2}+27a_{1}^{2}-9a_{1}a_{2}a_{3}+2a_{2}^{3}.
\]

In the transfer matrix technique in the thermodynamic limit ($N\to\infty$),
the partition function (\ref{eq:PF}) is 
\[
Z=\lambda_{1}^{N},
\]
where $\lambda_{1}$ is the principal eigenvalue of the transfer matrix,
which for this type of matrix always exists by the Perron--Frobenius
theorem~\citep{Domb:1960,Horn:2013}.

As a result, all thermodynamic functions of the system, including
the Helmholtz free energy per spin,
\[
F=-\frac{T}{N}\ln Z=-T\ln\lambda_{1},
\]
entropy
\begin{equation}
S=-\frac{\partial F}{\partial T}=\ln\lambda_{1}+\frac{T}{\lambda_{1}}\frac{\partial\lambda_{1}}{\partial T},\label{eq:S0}
\end{equation}
and heat capacity
\begin{equation}
C=-T\frac{\partial^{2}F}{\partial T^{2}}=2\frac{T}{\lambda_{1}}\frac{\partial\lambda_{1}}{\partial T}+\frac{T^{2}}{\lambda_{1}}\frac{\partial^{2}\lambda_{1}}{\partial T^{2}}-\frac{T^{2}}{\lambda_{1}^{2}}\left(\frac{\partial\lambda_{1}}{\partial T}\right)^{2}\label{eq:CV}
\end{equation}
are defined only in terms of the principal eigenvalue of the transfer
matrix~\citep{Baxter:1982,Nolting:2009,Mussardo:2010,Gould:2010}.

At the end of this section, it should be said that we know only one
paper \citep{Martinez-Garcilazo:2009}, in which, for the one-dimensional
Ising model, taking into account the interaction of spins at the sites
of third neighbors, a transfer matrix is constructed and expressions
for its eigenvalues are obtained in an explicit form. Unfortunately,
the results described in \citep{Martinez-Garcilazo:2009} are not
correct. This conclusion can be easily verified by obtaining an expression
from the characteristic equation presented in \citep{Martinez-Garcilazo:2009}
in the particular cases of smaller number of interactions in the model.

Taking into account the interaction only between the spins at the
nearest neighbors in the chain, that is, when
\[
K_{1}\neq0,\quad K_{2}=0,\quad K_{3}=0,
\]
we must get the well-known result of \citep{Ising:1925} for the characteristic
equation that contains the principal eigenvalue, in the form of
\[
\lambda-2\cosh K_{1}=0,
\]
but from the paper \citep{Martinez-Garcilazo:2009}, the characteristic
equation is transformed to the form
\[
\lambda-(2\cosh K_{1})^{3}=0.
\]

Also, when taking into account the interaction between the spins at
the first and second neighbors in the chain, where
\[
K_{1}\neq0,\quad K_{2}\neq0,\quad K_{3}=0,
\]
the characteristic equation that contains the principal eigenvalue,
must have the form
\[
\lambda^{2}-2\lambda e^{K_{2}}\cosh K_{1}+2\sinh(2K_{2})=0,
\]
as shown in Refs.~\citep{Oguchi:1965,Stephenson:1970}, but in the
paper \citep{Martinez-Garcilazo:2009} the equation reduces to quite
different form
\[
\lambda^{2}-2\lambda e^{-K_{2}}[3\cosh K_{1}+e^{4K_{2}}\cosh(3K_{1})]+8\sinh^{3}(2K_{2})=0.
\]
Obviously, the results of paper \citep{Martinez-Garcilazo:2009} are
not correct.

Note that the characteristic equation obtained in our work (\ref{eq:N3:CP1}),
in both particular cases of the number of interactions in the system,
gives the correct results.

\section{Magnetic phase diagram of the ground state of the system}

The model contains only eight variants of the relationship of the
parameters of the exchange interactions between the spins at the sites
of the first, second, and third neighbors of the chain. These relations
are
\begin{equation}
(J_{1}<0,J_{2}>0,J_{3}<0),\quad(J_{1}>0,J_{2}>0,J_{3}>0),\label{eq:N3:Q:41}
\end{equation}
\begin{equation}
(J_{1}<0,J_{2}>0,J_{3}>0),\quad(J_{1}>0,J_{2}>0,J_{3}<0),\label{eq:N3:Q:14}
\end{equation}
\begin{equation}
(J_{1}<0,J_{2}<0,J_{3}<0),\quad(J_{1}>0,J_{2}<0,J_{3}>0),\label{eq:N3:Q:32}
\end{equation}
\begin{equation}
(J_{1}<0,J_{2}<0,J_{3}>0),\quad(J_{1}>0,J_{2}<0,J_{3}<0).\label{eq:N3:Q:23}
\end{equation}
The first two sets (\ref{eq:N3:Q:41}) correspond to the aggravated
antiferromagnetic and ferromagnetic types of exchange interactions.
The last six sets of the parameters (\ref{eq:N3:Q:14})--(\ref{eq:N3:Q:23})
define the system with competing exchange interactions between spins.

The magnetic phase diagram of the ground state of the model is determined
by the behavior of the minimum energy of the spin system configurations
at zero temperature, depending on the model parameters
\begin{equation}
E_{0}=\min\{E\},\label{eq:E:0}
\end{equation}
where the configuration energy itself is an internal energy
\[
U=-T^{2}\frac{\partial}{\partial T}\frac{F}{T}=\frac{T^{2}}{\lambda_{1}}\frac{\partial\lambda_{1}}{\partial T},
\]
per lattice site at zero temperature
\[
E=\lim_{T\to0}U,
\]
which is explicitly specified by the operator of the total energy
of the system~(\ref{eq:H:0}) and is found from the function
\begin{equation}
E=-\frac{1}{m}\sum_{i=1}^{m}\sum_{p=1}^{b}J_{p}\frac{\sigma_{i+b-p}\sigma_{i+b}+\sigma_{i+b}\sigma_{i+b+p}}{2},\label{eq:N3:E}
\end{equation}
where $m$ is the number of sites in the configuration, $b$ is the
number of exchange interactions of the chain spins in the problem
($b=3$), $J_{p}$ is the parameter of the exchange interaction between
lattice spins at neighboring sites of the $p$-level.

Only five types of spin configurations with minimal energy are realized
in the ground state of the system, depending on the signs of the parameters
of the exchange interactions of the chain spins.

The first type of spin configurations is characterized by antiferromagnetic
ordering, which corresponds to a set
\begin{equation}
C_{\text{A}2}=\left\{ \begin{array}{ccccccc}
\uparrow & \downarrow & \uparrow & \downarrow & \uparrow & \downarrow & \cdots\\
\downarrow & \uparrow & \downarrow & \uparrow & \downarrow & \uparrow & \cdots
\end{array}\right\} ,\label{eq:C:A2}
\end{equation}
consisting of two sequences (with alternating spin projections along
and against the direction of the $z$-axis) with equal energies
\begin{equation}
E_{\text{A}2}=J_{1}-J_{2}+J_{3}.\label{eq:N3:E:A2}
\end{equation}
For this configuration, we introduce the designation A2 \citep{Zarubin:2019:}.

The second type of spin configurations is characterized by magnetic
ordering with a tripling of the translation period (configuration
designation~A3),
\begin{equation}
C_{\text{A}3}=\left\{ \begin{array}{ccccccc}
\uparrow & \uparrow & \downarrow & \uparrow & \uparrow & \downarrow & \cdots\\
\uparrow & \downarrow & \uparrow & \uparrow & \downarrow & \uparrow & \cdots\\
\downarrow & \uparrow & \uparrow & \downarrow & \uparrow & \uparrow & \cdots\\
\downarrow & \downarrow & \uparrow & \downarrow & \downarrow & \uparrow & \cdots\\
\downarrow & \uparrow & \downarrow & \downarrow & \uparrow & \downarrow & \cdots\\
\uparrow & \downarrow & \downarrow & \uparrow & \downarrow & \downarrow & \cdots
\end{array}\right\} ,\label{eq:C:A3}
\end{equation}
which consists of six configurations with equal energies
\begin{equation}
E_{\text{A}3}=\frac{J_{1}+J_{2}-3J_{3}}{3}.\label{eq:N3:E:A3}
\end{equation}

The third type is determined by magnetic ordering with quadruple period
(configuration designation~A4),
\begin{equation}
C_{\text{A}4}=\left\{ \begin{array}{ccccccc}
\uparrow & \uparrow & \downarrow & \downarrow & \uparrow & \uparrow & \cdots\\
\uparrow & \downarrow & \downarrow & \uparrow & \uparrow & \downarrow & \cdots\\
\downarrow & \uparrow & \uparrow & \downarrow & \downarrow & \uparrow & \cdots\\
\downarrow & \downarrow & \uparrow & \uparrow & \downarrow & \downarrow & \cdots
\end{array}\right\} ,\label{eq:C:A4}
\end{equation}
which consists of four configurations with equal energies
\begin{equation}
E_{\text{A}4}=J_{2}.\label{eq:N3:E:A4}
\end{equation}

The fourth type is characterized by magnetic ordering with sextuple
period (configuration designation~A6),
\begin{equation}
C_{\text{A}6}=\left\{ \begin{array}{ccccccc}
\uparrow & \uparrow & \uparrow & \downarrow & \downarrow & \downarrow & \cdots\\
\uparrow & \uparrow & \downarrow & \downarrow & \downarrow & \uparrow & \cdots\\
\uparrow & \downarrow & \downarrow & \downarrow & \uparrow & \uparrow & \cdots\\
\downarrow & \downarrow & \downarrow & \uparrow & \uparrow & \uparrow & \cdots\\
\downarrow & \downarrow & \uparrow & \uparrow & \uparrow & \downarrow & \cdots\\
\downarrow & \uparrow & \uparrow & \uparrow & \downarrow & \downarrow & \cdots
\end{array}\right\} ,\label{eq:C:A6}
\end{equation}
which consists of six configurations with equal energies
\begin{equation}
E_{\text{A}6}=\frac{-J_{1}+J_{2}+3J_{3}}{3}.\label{eq:N3:E:A6}
\end{equation}

The fifth type of spin configurations is characterized by ferromagnetic
ordering (configuration designation~F2) with a set
\begin{equation}
C_{\text{F}2}=\left\{ \begin{array}{ccccccc}
\uparrow & \uparrow & \uparrow & \uparrow & \uparrow & \uparrow & \cdots\\
\downarrow & \downarrow & \downarrow & \downarrow & \downarrow & \downarrow & \cdots
\end{array}\right\} ,\label{eq:C:F2}
\end{equation}
which consists of two sequences (along and against the direction of
the $z$-axis) with equal energies
\begin{equation}
E_{\text{F}2}=-(J_{1}+J_{2}+J_{3}).\label{eq:N3:E:F2}
\end{equation}

Other types of magnetic ordering, i.e. spin configurations with quintuple,
septuple, or higher increase in the translation period, do not have
the minimum ground state energy at any ratios of the exchange parameters
of the system.

Recall that the above configurations of the ground state correspond
to the following designations~$\langle1\rangle$, $\langle12\rangle$,
$\langle2\rangle$, $\langle3\rangle$, $\langle\infty\rangle$ introduced
in \citep{Fisher:1980,Fisher:1981} and widely used in the ANNNI model
\citep{Selke:1985,Selke:1988,Yeomans:1988}.

\begin{figure*}
\centering \includegraphics{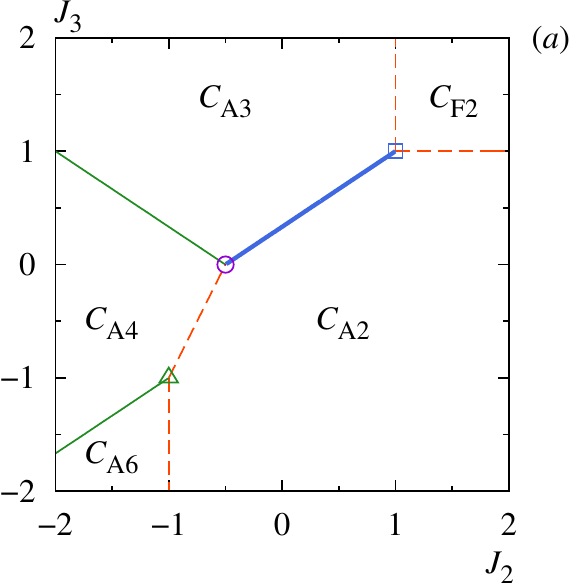}\hfill{}\includegraphics{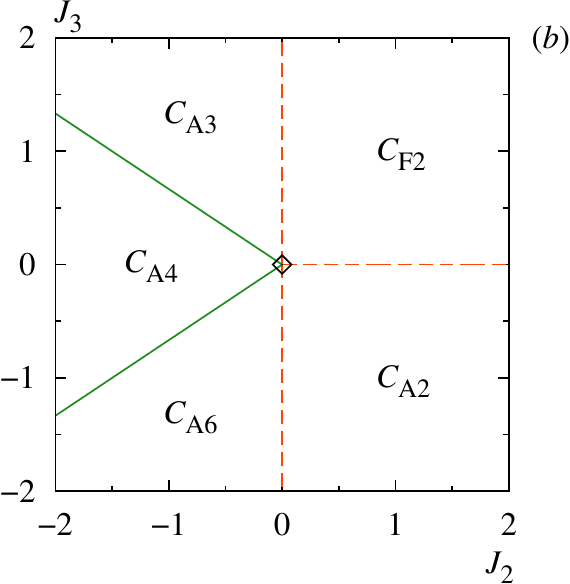}\hfill{}\includegraphics{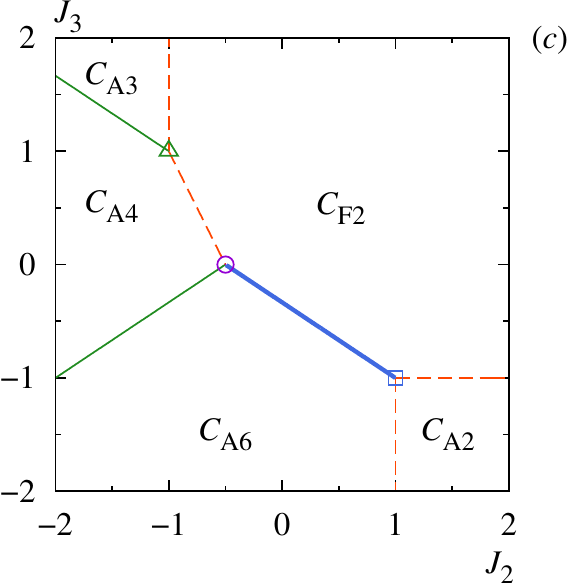}
\protect\caption{Magnetic phase diagram of the ground state of the Ising chain with
the exchange interactions of spins at the sites of the linear chain
of the first, second, and third neighbors with (a) antiferromagnetic
($J_{1}=-1$), (b) missing ($J_{1}=0$) and (c) ferromagnetic ($J_{1}=+1$)
interactions between the spins at the sites of nearest neighbors}
\label{fig:N3:PD}
\end{figure*}

Thus, the spin configurations under consideration have corresponding
energies in the following ranges of interaction parameters
\[
E_{0}=\begin{cases}
E_{\text{A}2}, & J_{1}\leqslant J_{2}\land J_{1}\leqslant2J_{2}-J_{3}\\
 & \land J_{1}\leqslant2J_{2}-3J_{3}\land J_{1}\leqslant-J_{3},\\
E_{\text{A}3},\quad & J_{1}\leqslant-J_{2}\land J_{1}\geqslant2J_{2}-3J_{3}\\
 & \land J_{1}\leqslant2J_{2}+3J_{3},\\
E_{\text{A}4}, & J_{1}\leqslant-2J_{2}+3J_{3}\land J_{1}\geqslant2J_{2}-J_{3}\\
 & \land J_{1}\geqslant2J_{2}+3J_{3}\land J_{1}\leqslant-2J_{2}-J_{3},\\
E_{\text{A}6}, & J_{1}\geqslant J_{2}\land J_{1}\geqslant-2J_{2}+3J_{3}\\
 & \land J_{1}\leqslant-2J_{2}-3J_{3},\\
E_{\text{F}2}, & J_{1}\geqslant-J_{2}\land J_{1}\geqslant-J_{3}\\
 & \land J_{1}\geqslant-2J_{2}-J_{3}\\
 & \land J_{1}\geqslant-2J_{2}-3J_{3}.
\end{cases}
\]
From this expression, we can obtain the ratios of the exchange parameters
of the model at which the rearrangement of the ordering structure
of the spin configurations of the ground state occurs,
\[
J_{1}=\begin{cases}
-J_{3}, & J_{2}>J_{3}\land J_{2}>-J_{3},\\
2J_{2}-3J_{3}, & J_{2}\leqslant J_{3}\land J_{3}>0,\\
2J_{2}-J_{3}, & J_{2}<J_{3}\land J_{3}<0,\\
J_{2}, & J_{2}>J_{3}\land J_{2}<-J_{3},\\
-J_{2}, & J_{2}<J_{3}\land J_{2}>-J_{3},\\
2J_{2}+3J_{3}, & J_{2}\leqslant-J_{3}\land J_{3}>0,\\
-2J_{2}+3J_{3},\quad & J_{2}\leqslant J_{3}\land J_{3}<0,\\
-2J_{2}-J_{3}, & J_{2}<-J_{3}\land J_{3}>0,\\
-2J_{2}-3J_{3}, & J_{2}\leqslant-J_{3}\land J_{3}<0,
\end{cases}
\]
with the formation of a complicated structure of the boundaries of
the regions of these configurations, as shown in the magnetic phase
diagram of the spin system \citep{Price:1983,Barreto:1985,Selke:1985},
and presented in Fig.~\ref{fig:N3:PD}.

Note that the structure of the magnetic phase diagram (see Fig.~\ref{fig:N3:PD}a
and \ref{fig:N3:PD}c) is antisymmetric with respect to the replacement
\begin{equation}
\{J_{1},J_{3}\}\Leftrightarrow\{-J_{1},-J_{3}\},\label{eq:J1:J3}
\end{equation}
which, upon further analysis of the model, allows one to consider
the behavior of thermodynamic quantities with only one sign of the
parameter of the exchange interaction between spins at the sites of
the nearest neighbors and at the same time fully describe the thermodynamics
of the system.

In the magnetic phase diagram, with the antiferromagnetic spin exchange
parameter at the sites of the second neighbors ($J_{2}<0$) at zero
temperature, a point is formed at the following ratios of the model
parameters
\begin{equation}
J_{2}=-|J_{1}|/2,\quad J_{3}=0,\label{eq:N3:PD:P0}
\end{equation}
which delimits three regions of spin configurations $C_{\text{A}4}$,
$C_{\text{A}3}$ and $C_{\text{A}2}$ for the antiferromagnetic parameter
of the exchange interaction between the spins at the sites of the
first neighbors ($J_{1}<0$), or the regions of $C_{\text{A}4}$,
$C_{\text{A}6}$ and $C_{\text{F}2}$ with the ferromagnetic parameter
($J_{1}>0$). This triple point (\ref{eq:N3:PD:P0}) in the phase
diagram (Fig.~\ref{fig:N3:PD}) is marked by a circlet and has the
ground state energy equal to $E=-|J_{1}|/2$.

In the case of a zero value of the spin exchange parameter at the
sites of the first neighbors ($J_{1}=0$), the triple point in the
phase diagram (\ref{eq:N3:PD:P0}) shifts to the position where the
values of all exchange parameters are zero,
\begin{equation}
J_{1}=J_{2}=J_{3}=0.\label{eq:N3:PD:P0:0}
\end{equation}
This point in the magnetic phase diagram (Fig.~\ref{fig:N3:PD}b)
is indicated by a rhombus, it is already common for the five spin
configuration regions of $C_{\text{A}2}$, $C_{\text{A}6}$, $C_{\text{A}4}$,
$C_{\text{A}3}$, and $C_{\text{F}2}$, and has the energy equal to
zero, $E=0$.

Note that this fact does not contradict the Gibbs phase rule, since
the considered spin system at a given point (\ref{eq:N3:PD:P0:0})
is defined as an Ising paramagnet in an absolutely frustrated state
\citep{Zarubin:2019:}.

Depending on the signs of the parameters of the exchange interactions
of spins at the sites of the first and third neighbors at the point
(\ref{eq:N3:PD:P0}), the following boundaries of the regions of spin
configurations are formed.

In the antiferro-antiferro-antiferromagnetic variant of the parameters
of the exchange interactions between the spins at the sites of the
nearest, second and third neighbors ($J_{1}<0$, $J_{2}<0$, $J_{3}<0$)
and with an increase of the parameter $|J_{3}|$, in the phase diagram
from the point (\ref{eq:N3:PD:P0}) a linear segment is formed
\begin{equation}
J_{3}=-J_{1}+2J_{2},\quad J_{1}<J_{2}<J_{1}/2,\quad J_{1}<J_{3}<0,\label{eq:N3:PD:L3}
\end{equation}
defining the common boundary of the configuration regions $C_{\text{A}4}$
and $C_{\text{A}2}$, which has the energy $E=J_{2}$.

Also, with an increase of the parameter $|J_{3}|$, and in the ferro-antiferro-ferromagnetic
variant of the exchange interaction parameters ($J_{1}>0$, $J_{2}<0$,
$J_{3}>0$), from the point (\ref{eq:N3:PD:P0}) a line segment is
formed
\begin{equation}
J_{3}=-J_{1}-2J_{2},\quad-J_{1}<J_{2}<-J_{1}/2,\quad0<J_{3}<J_{1},\label{eq:N3:PD:L8}
\end{equation}
defining the boundary of the configuration regions $C_{\text{A}4}$
and $C_{\text{F}2}$, which has an energy $E=J_{2}$.

The considered segments of the boundaries of the phase diagram (\ref{eq:N3:PD:L3})
and (\ref{eq:N3:PD:L8}) start at the point (\ref{eq:N3:PD:P0}),
and with a further increase of the parameter $|J_{3}|$ end in position
\begin{equation}
J_{2}=-|J_{1}|,\quad J_{3}=J_{1},\label{eq:N3:PD:P1}
\end{equation}
in which there arises another spin configuration $C_{\text{A}6}$
with the antiferromagnetic parameter of the exchange interaction between
the spins at the sites of the first neighbors ($J_{1}<0$) or the
configuration $C_{\text{A}3}$ with the ferromagnetic parameter ($J_{1}>0$).
The triple point (\ref{eq:N3:PD:P1}) in the phase diagram (Fig.~\ref{fig:N3:PD})
is marked by a triangle and has the energy $E=-|J_{1}|$.

With a further increase of the parameter $|J_{3}|$ in the phase diagram
at the point (\ref{eq:N3:PD:P1}) the boundaries of already three
regions of spin configurations are formed.

Thus, in the antiferro-antiferro-antiferromagnetic variant of the
parameters of the exchange interactions between the spins at the sites
of the nearest, second, and third neighbors ($J_{1}<0$, $J_{2}<0$,
$J_{3}<0$), the lines emerging from the point (\ref{eq:N3:PD:P1})
defining the boundaries of the spin configurations $C_{\text{A}4}$,
$C_{\text{A}6}$ and $C_{\text{A}2}$ are given by
\begin{equation}
J_{3}=\frac{J_{1}+2J_{2}}{3},\quad J_{2}\leqslant J_{1},\quad J_{3}\leqslant J_{1},\label{eq:N3:PD:L7}
\end{equation}
for regions $C_{\text{A}4}$ and $C_{\text{A}6}$ with the energy
of states at the boundary $E=J_{2}$, and also
\begin{equation}
J_{2}=J_{1},\quad J_{3}<-|J_{1}|\label{eq:N3:PD:L4}
\end{equation}
for the boundary between the configurations $C_{\text{A}6}$ and $C_{\text{A}2}$
with the energy $E=J_{3}$.

In the ferro-antiferro-ferromagnetic variant of the exchange interaction
parameters ($J_{1}>0$, $J_{2}<0$, $J_{3}>0$), the lines emerging
from the point (\ref{eq:N3:PD:P1}) defining the boundaries of the
spin configurations $C_{\text{A}4}$, $C_{\text{A}3}$ and $C_{\text{F}2}$
are given by the expressions
\begin{equation}
J_{3}=\frac{J_{1}-2J_{2}}{3},\quad J_{2}<J_{1}/2,\label{eq:N3:PD:L6}
\end{equation}
\begin{equation}
J_{2}=-J_{1},\quad J_{3}>|J_{1}|,\label{eq:N3:PD:L5}
\end{equation}
and the energies of states on these lines are respectively equal to
$E=J_{2}$ and $E=-J_{3}$.

On the other hand, in the antiferro-antiferro-ferromagnetic variant
of the exchange interaction parameters ($J_{1}<0$, $J_{2}<0$, $J_{3}>0$)
with increasing parameter $|J_{3}|$ at the point (\ref{eq:N3:PD:P0})
in the phase diagram the segments
\begin{equation}
J_{3}=\frac{J_{1}-2J_{2}}{3},\quad J_{2}\leqslant-J_{1},\quad J_{3}\geqslant J_{1},\label{eq:N3:PD:L6:}
\end{equation}
are formed, defining the corresponding boundaries of the configuration
regions $C_{\text{A}4}$ and $C_{\text{A}3}$ with the ground state
energy equal to $E=J_{2}$, and also the segments
\begin{equation}
J_{3}=-\frac{J_{1}-2J_{2}}{3},\quad J_{1}/2<J_{2}\leqslant-J_{1},\quad0<J_{3}\leqslant-J_{1},\label{eq:N3:PD:L2}
\end{equation}
are formed, defining the boundaries of the configuration regions $C_{\text{A}3}$
and $C_{\text{A}2}$ with the energy
\begin{equation}
E=\frac{J_{1}-J_{3}}{2}.\label{eq:N3:E:L2}
\end{equation}

In the ferro-antiferro-antiferromagnetic variant of the exchange interaction
parameters ($J_{1}>0$, $J_{2}<0$, $J_{3}<0$), the boundaries are
formed at the point (\ref{eq:N3:PD:P0}), which are described by the
expressions
\begin{equation}
J_{3}=\frac{J_{1}+2J_{2}}{3},\quad J_{2}<-J_{1}/2\label{eq:N3:PD:L7:}
\end{equation}
for regions of spin configurations $C_{\text{A}4}$ and $C_{\text{A}6}$,
and
\begin{equation}
J_{3}=-\frac{J_{1}+2J_{2}}{3},\quad-J_{1}/2<J_{2}\leqslant J_{1},\quad-J_{1}\leqslant J_{3}<0,\label{eq:N3:PD:L9}
\end{equation}
for regions $C_{\text{A}6}$ and $C_{\text{F}2}$. The energies of
the system at zero temperature on the lines (\ref{eq:N3:PD:L7:})
and (\ref{eq:N3:PD:L9}) are respectively equal to $E=J_{2}$ and
\[
E=-\frac{J_{1}-J_{3}}{2}.
\]

Depending on the sign of the exchange interaction parameters between
the spins at the sites of the nearest neighbors, the segments (\ref{eq:N3:PD:L2})
or (\ref{eq:N3:PD:L9}) determine the boundaries in the phase diagram,
which, passing through the position
\begin{equation}
J_{2}=0,\quad J_{3}=-J_{1}/3,\label{eq:N3:PD:P3}
\end{equation}
connect the point (\ref{eq:N3:PD:P0}) with the point
\begin{equation}
J_{2}=|J_{1}|,\quad J_{3}=-J_{1}.\label{eq:N3:PD:P2}
\end{equation}
The triple point (\ref{eq:N3:PD:P2}) in the phase diagram (Fig.~\ref{fig:N3:PD})
is marked by a square and has the energy $E=-|J_{1}|$.

In the antiferro-ferro-ferromagnetic variant of the exchange interaction
parameters ($J_{1}<0$, $J_{2}>0$, $J_{3}>0$) and with increasing
parameter $|J_{3}|$ at the triple point in the magnetic phase diagram
(\ref{eq:N3:PD:P2}), another configuration region $C_{\text{F}2}$
is formed, the boundaries of which with the corresponding regions
of the spin configurations $C_{\text{A}3}$ and$C_{\text{A}2}$ are
defined by the following linear laws (\ref{eq:N3:PD:L5}) and
\begin{equation}
J_{2}>|J_{1}|,\quad J_{3}=-J_{1}.\label{eq:N3:PD:L1}
\end{equation}
The energies of the ground state of the system in the positions (\ref{eq:N3:PD:L5})
and (\ref{eq:N3:PD:L1}) are respectively equal to $E=-J_{3}$ and
$E=-J_{2}$.

On the other hand, in the ferro-ferro-antiferromagnetic variant of
the exchange interactions parameters ($J_{1}>0$, $J_{2}>0$, $J_{3}<0$)
and with increasing parameter $|J_{3}|$ at the triple point in the
magnetic phase diagram (\ref{eq:N3:PD:P2}), another configuration
region $C_{\text{A}2}$ is formed, the boundaries of which with the
corresponding regions of the spin configurations $C_{\text{A}6}$
and $C_{\text{F}2}$ are determined by the linear laws (\ref{eq:N3:PD:L4})
and (\ref{eq:N3:PD:L1}) with the corresponding energies $E=J_{3}$
and $E=-J_{2}$.

Note that in the case of zero exchange interaction between the spins
at the sites of the nearest neighbors ($J_{1}=0$), the above segments
of the boundary between the configurations $C_{\text{A}4}$ and $C_{\text{F}2}$
(\ref{eq:N3:PD:L3}); $C_{\text{A}4}$ and $C_{\text{F}2}$ (\ref{eq:N3:PD:L8}),
as well as $C_{\text{A}3}$ and $C_{\text{F}2}$ (\ref{eq:N3:PD:L2});
$C_{\text{A}6}$ and $C_{\text{F}2}$ (\ref{eq:N3:PD:L9}) are absent,
i.e. all the triple points described above are combined where the
quintuple point is formed, as one can see in Fig.~\ref{fig:N3:PD}b.

Thus, the lines in the magnetic phase diagram of the ground state
demonstrate the boundaries of the regions of spin configurations on
which a qualitative change in the structure of magnetic ordering of
the ground state occurs.

In Fig.~\ref{fig:N3:PD} dashed lines indicate the boundaries (determined
by the following relations of the model parameters (\ref{eq:N3:PD:L1}),
(\ref{eq:N3:PD:L5}), (\ref{eq:N3:PD:L3}) and (\ref{eq:N3:PD:L4})
with the antiferromagnetic parameter of the exchange interaction between
the spins at the sites of the first neighbors ($J_{1}<0$), as well
as (\ref{eq:N3:PD:L1}), (\ref{eq:N3:PD:L4}), (\ref{eq:N3:PD:L8})
and (\ref{eq:N3:PD:L5}) with the ferromagnetic exchange interaction
parameter ($J_{1}>0$)), on which the rearrangement of the ground
state ordering occurs, and the number of configurations of the system
with minimum energy is equal to the sum of the configurations of the
regions adjacent to the boundary.

The solid lines in Fig.~\ref{fig:N3:PD} indicate the boundaries
(determined by the following relations of the model parameters (\ref{eq:N3:PD:L7}),
(\ref{eq:N3:PD:L6}) and (\ref{eq:N3:PD:L2}) with the antiferromagnetic
parameter ($J_{1}<0$), as well as (\ref{eq:N3:PD:L6}), (\ref{eq:N3:PD:L7})
and (\ref{eq:N3:PD:L9}) with the ferromagnetic parameter ($J_{1}>0$)),
on which the number of configurations of the system with the minimum
energy is greater than the sum of the configurations of the adjacent
regions of the phase diagram.

Such a multitude of spin configurations of the system at zero temperature
is associated with the rearrangement of the magnetic structure and
the appearance at the given phase a point (in the thermodynamic limit)
of an infinite number of spin configurations, including a violation
of translational invariance.

\begin{figure}[ht]
\centering \includegraphics[scale=1.1]{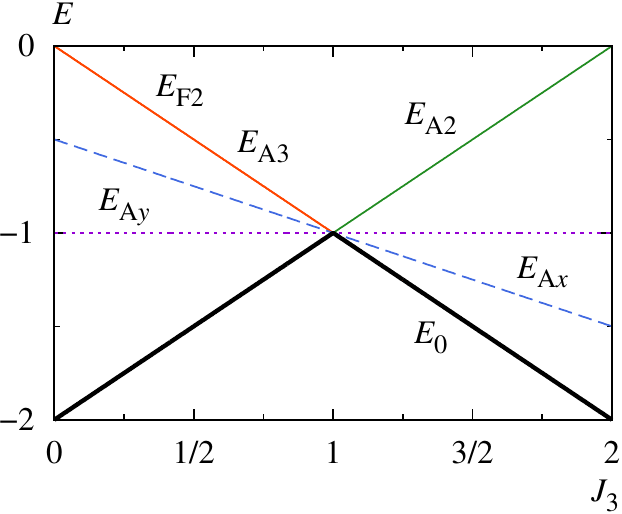}
\protect\caption{The energies of the ground state of spin configurations depending
on the ferromagnetic exchange interaction parameter between spins
at the sites of third neighbors ($J_{3}\geqslant0$) in the magnetic
phase space region, where $J_{1}=-1$ and $J_{2}=+1$. The frustration
regime in the system occurs at the point $J_{3}=+1$. The solid lines
indicate the energies $E_{\text{A}2}$ (\ref{eq:N3:E:A2}) and $E_{\text{A}3}$
(\ref{eq:N3:E:A3}), the dashed line indicates $E_{\text{A}x}$ (\ref{eq:N3:E:Ax}),
the dotted line indicates $E_{\text{A}y}$ (\ref{eq:N3:E:Ay}), and
the thick solid line indicates the minimum energy $E_{0}$ (\ref{eq:E:0})}
\label{fig:N3:CNF3:P03} 
\end{figure}

This situation can be demonstrated by the following example. Note
that in the phase diagram shown in Fig.~\ref{fig:N3:PD}a, the boundary
indicated by the solid line (\ref{eq:N3:PD:L2}), with the energy
(\ref{eq:N3:E:L2}{]}) corresponds to the adjacent regions of the
spin configurations $C_{\text{A}2}$ and $C_{\text{A}3}$ with the
corresponding energies (\ref{eq:N3:E:A2}) and (\ref{eq:N3:E:A3}).
In the antiferro-ferro-ferromagnetic variant of the exchange interaction
parameters, there are other configuration sequences, for example,
\[
\left\{ \begin{array}{ccccccccc}
\uparrow & \downarrow & \uparrow & \downarrow & \downarrow & \uparrow & \downarrow & \uparrow & \cdots\\
\downarrow & \uparrow & \downarrow & \downarrow & \uparrow & \downarrow & \downarrow & \uparrow & \cdots
\end{array}\right\} ,
\]
with the same energy
\begin{equation}
E_{\text{A}x}=\frac{J_{1}-J_{3}}{2}\label{eq:N3:E:Ax}
\end{equation}
or
\[
\left\{ \begin{array}{ccccccccccccc}
\uparrow & \uparrow & \downarrow & \uparrow & \downarrow & \uparrow & \uparrow & \downarrow & \uparrow & \downarrow & \uparrow & \downarrow & \cdots\\
\uparrow & \downarrow & \downarrow & \uparrow & \downarrow & \uparrow & \uparrow & \downarrow & \uparrow & \downarrow & \uparrow & \downarrow & \cdots\\
\downarrow & \uparrow & \downarrow & \uparrow & \uparrow & \downarrow & \uparrow & \uparrow & \downarrow & \uparrow & \downarrow & \uparrow & \cdots\\
\downarrow & \downarrow & \uparrow & \downarrow & \uparrow & \uparrow & \downarrow & \uparrow & \downarrow & \uparrow & \downarrow & \uparrow & \cdots
\end{array}\right\} 
\]
with the energy
\begin{equation}
E_{\text{A}y}=\frac{2J_{1}-J_{2}}{3},\label{eq:N3:E:Ay}
\end{equation}
which also have minimal ground state energy only at the phase boundary
of the regions under consideration (\ref{eq:N3:E:L2}), as shown in
Fig.~\ref{fig:N3:CNF3:P03}.

In the terminology of the works \citep{Fisher:1981,Pokrovskii:1982:,Selke:1985,Barreto:1985,Yeomans:1987,Yeomans:1988},
in the magnetic phase diagram of the ground state (Fig.~\ref{fig:N3:PD}),
the triple points described above are called mutiphase points, and
solid lines are called mutiphase lines.

\section{Residual entropy of the system at zero temperature}

In the magnetic phase diagram of the ground state of the model with
competing exchange interactions of spins at the sites of the first,
second, and third neighbors (Fig.~\ref{fig:N3:PD}) in the regions
outside the boundaries of the spin configurations and at the boundaries
marked by the dashed lines, the corresponding values of the zero-temperature
(residual) entropy are zero,
\begin{equation}
\lim_{T\to0}S=0,\label{eq:S0:0}
\end{equation}
and at the boundaries indicated by solid lines, the entropy of the
ground state is no longer zero,
\begin{equation}
\lim_{T\to0}S>0.\label{eq:S0:fr}
\end{equation}

This result (\ref{eq:S0:fr}) does not contradict the third law of
thermodynamics, since the entropy is determined 
\begin{equation}
S-S_{0}=\intop_{T_{0}}^{T}\frac{\delta Q}{T}\label{eq:DS}
\end{equation}
up to the integration constant $S_{0}\geqslant0$. This constant is
chosen equal to zero ($S_{0}=0$) only in the formulation of the Nernst--Planck
theorem for equilibrium systems with nondegenerate ground state \citep{Sommerfeld:1956}.

On the other hand, if the Gibbs entropy of the ground state of the
system is greater than zero,
\[
S(T=0)=\ln W>0,
\]
this suggests that the system experiences degeneracy of the ground
state, since the statistical weight characterizing the multiplicity
of degeneracy of the system is greater than unity ($W>1$) \citep{Nolting8:2018}.

Thus, the state of a system in which the entropy of the ground state
is not zero (\ref{eq:S0:fr}) should be called \emph{frustrated} \citep{Zarubin:2019:}.

It should be noted that, in contrast to the situation with a smaller
number of interactions in the Ising model \citep{Zarubin:2019:},
in the case under consideration there are much more relations of exchange
interaction parameters in the system, at which the frustrated system
behaves differently.

For example, in the magnetic phase diagram in the absence of exchange
between the spins of the chain (\ref{eq:N3:PD:P0:0}) at the quintuple
point marked by rhombus in Fig.~\ref{fig:N3:PD}b, a paramagnetic
state is realized, characterized by that all configurations of the
system have the same probability and have the same zero energy. The
entropy of such a state of the system is equal to the natural logarithm
of two,
\begin{equation}
S=\ln2\approx0.693,\label{eq:N3:S0:0}
\end{equation}
and is the same at any temperature. From this it is clear that \emph{the
Ising paramagnet is an absolutely frustrated system} \citep{Zarubin:2019:}.

With the ratio of the exchange interaction parameters of the model
(\ref{eq:N3:PD:P0}) at the triple point marked in the phase diagram
(Fig.~\ref{fig:N3:PD}) by a circlet, the entropy at zero temperature
is equal to the natural logarithm of the golden ratio,
\begin{equation}
\lim_{T\to0}S=\ln\frac{1+\sqrt{5}}{2}\approx0.481.\label{eq:N3:S0:1}
\end{equation}

In the phase diagram (Fig.~\ref{fig:N3:PD}), thick solid lines indicate
the boundaries of the regions of spin configurations with the ratios
of the exchange parameters (\ref{eq:N3:PD:L2}) and (\ref{eq:N3:PD:L9}),
as well as square marks indicate triple points (\ref{eq:N3:PD:P2}),
at which the residual entropy is equal to
\begin{equation}
\lim_{T\to0}S=\ln\left[\frac{1}{3}\left(1+\vartheta_{2}+\frac{1}{\vartheta_{2}}\right)\right]\approx0.382,\label{eq:N3:S0:2}
\end{equation}
\[
\vartheta_{2}=\sqrt[3]{\frac{3^{3}+2+\sqrt{(3^{3}+2^{2})3^{3}}}{2}}.
\]
Also, thin solid lines mark the boundaries of the configurations when
the ratio of the model parameters (\ref{eq:N3:PD:L7}) and (\ref{eq:N3:PD:L6}),
as well as triangular points indicate the triple points (\ref{eq:N3:PD:P1})
in the phase diagram (Fig.~\ref{fig:N3:PD}), at which the residual
entropy has the following value
\begin{equation}
\lim_{T\to0}S=\ln\left[\frac{1}{3}\left(\vartheta_{3}+\frac{3}{\vartheta_{3}}\right)\right]\approx0.281,\label{eq:N3:S0:3}
\end{equation}
\[
\vartheta_{3}=\sqrt[3]{\frac{3^{3}+\sqrt{(3^{3}-2^{2})3^{3}}}{2}}.
\]
Note that the positions of the frustration of the system in the phase
diagram correspond to multiphase points and lines in \citep{Barreto:1985,Selke:1985}.

It should be noted, that the expressions presented above for zero-temperature
entropy can be written as
\[
S=\ln x_{\text{max}},
\]
where the argument of the natural logarithm is the statistical weight
of the system, which is defined as the maximum real root of the corresponding
equation. Thus for residual entropy (\ref{eq:N3:S0:1}) this equation
has the form
\[
x^{2}-x-1=0,
\]
for (\ref{eq:N3:S0:2}) is
\[
x^{3}-x^{2}-1=0,
\]
and for entropy (\ref{eq:N3:S0:3}) this equation is
\[
x^{3}-x-1=0.
\]
For the presented equations, there always exists a single positive
real root.

\section{Thermodynamics of the system in the frustration regime and its vicinity}

The thermodynamic functions of the model demonstrate a complicated
temperature behavior at various ratios of the parameters of the exchange
interaction of spins at the sites of the first, second, and third
neighbors.

At zero temperature, the entropy of the Ising chain has either a zero
value (\ref{eq:S0:0}) or a finite value in the frustration regime,
with values (\ref{eq:N3:S0:0}), (\ref{eq:N3:S0:1}), (\ref{eq:N3:S0:2})
or (\ref{eq:N3:S0:3}), as well as at an infinitely high temperature
(for any values of the exchange parameters of the model), the entropy
of the system is equal to the natural logarithm of two,
\begin{equation}
\lim_{T\to\infty}S=\ln2,\label{eq:S0:T8}
\end{equation}
where the statistical weight of the system ($W=2$) corresponds to
the number of states at the site in the model.

\begin{figure}[ht]
\centering \includegraphics[scale=1.1]{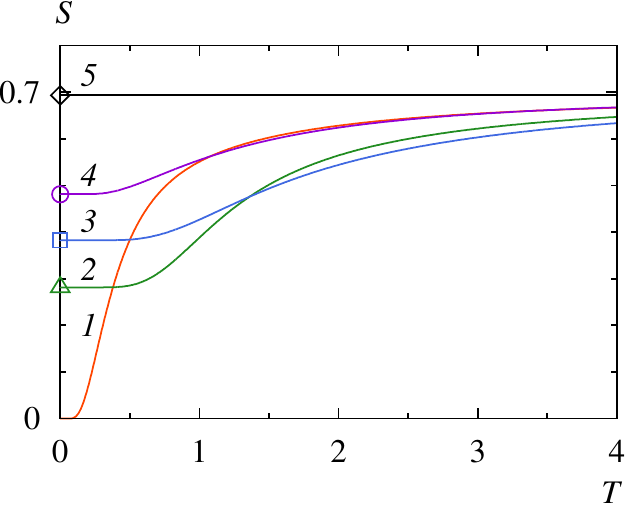}
\protect\caption{The temperature dependence of the entropy (\ref{eq:S0}) of the Ising chain at five points in the magnetic phase space, where the ratios
of the parameters of the exchange interactions of spins are ($J_{1}=-1$,
$J_{2}=-0.7$, $J_{3}=-0.4$) for line 1, ($J_{1}=J_{2}=J_{3}=-1$)
for line 2, ($J_{1}=-J_{2}=-J_{3}=-1$) for 3, ($J_{1}=-1$, $J_{2}=-1/2$,
$J_{3}=0$) for 4, and ($J_{1}=J_{2}=J_{3}=0$) for 5}
\label{fig:N3:S0:5} 
\end{figure}

Thus, when the temperature changes, the entropy of the system has
values in the range
\[
0\leqslant S\leqslant\ln2.
\]
Figure~\ref{fig:N3:S0:5} shows the entropy behavior of the model
outside the frustration regime (line 1), in the frustration regime
(lines 2, 3, 4), as well as in the completely frustrated system regime
(line 5). At an infinitely high temperature and any values of the
exchange interaction parameters of the system, the entropy tends to
a finite value equal to (\ref{eq:S0:T8}).

In turn, the heat capacity of the system is zero for any ratio of
the exchange parameters of the spins, and at zero and infinitely high
temperatures,
\begin{equation}
\lim_{T\to0}C=0,\quad\lim_{T\to\infty}C=0.\label{eq:CV:T08}
\end{equation}
At intermediate temperatures, the heat capacity has a maximum, which
splits into several peaks in the vicinity of the frustration point.
For various ratios of the model parameters, several types (scenarios)
of the behavior of the temperature evolution of heat capacity are
formed.

\begin{figure}[ht]
\centering \includegraphics[scale=1.1]{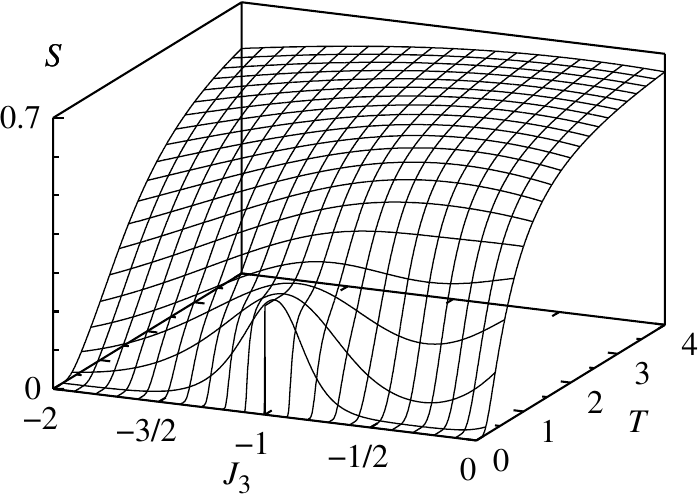}
\protect\caption{The temperature evolution of the entropy (\ref{eq:S0}) of the Ising chain in the antiferro-antiferro-antiferromagnetic variant of the
parameters of the exchange interactions of spins, where $J_{1}=J_{2}=-1$,
and $J_{3}<0$}
\label{fig:N3:S0:P1:mmm100} 
\end{figure}

\begin{figure}[ht]
\centering \includegraphics[scale=1.1]{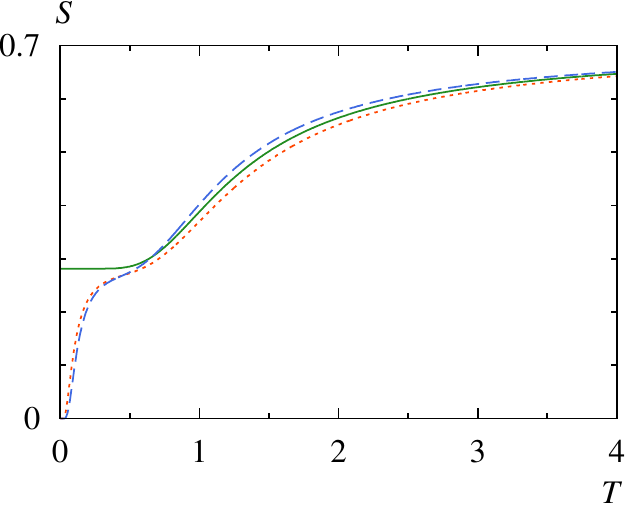}
\protect\caption{The temperature dependence of the entropy (\ref{eq:S0}) of the Ising chain in the vicinity of the frustration point (\ref{eq:N3:PD:P1})
of the system, where the parameters of the exchange interactions of
spins are equal to $J_{1}=J_{2}=-1$, and the values of $J_{3}=\{-1.1,-1,-0.9\}$
correspond to the dotted, solid, and dashed lines}
\label{fig:N3:S0:P1:mmm100:} 
\end{figure}

For example, in the case of the antiferro-antiferro-anti\-ferro\-magnetic
variant of the exchange interaction parameters of the model ($J_{1}<0$,
$J_{2}<0$, $J_{3}<0$) with the ratio of the quantities $J_{2}\leqslant J_{1}$
and $J_{3}\leqslant J_{1}$, frustration occurs in the system at a
certain ratio of the model parameters (\ref{eq:N3:PD:P1}), therefore,
only in this case, the residual entropy is not equal to zero (\ref{eq:N3:S0:3}),
and for all other ratios of the parameters in this interval of values,
the entropy of the ground state is zero (\ref{eq:S0:0}), as shown
in Fig.~\ref{fig:N3:S0:P1:mmm100}. The temperature dependence of
entropy at the point of frustration (\ref{eq:N3:PD:P1}) and its small
vicinity is presented in more detail in Fig.~\ref{fig:N3:S0:P1:mmm100:}.

\begin{figure}[ht]
\centering \includegraphics[scale=1.1]{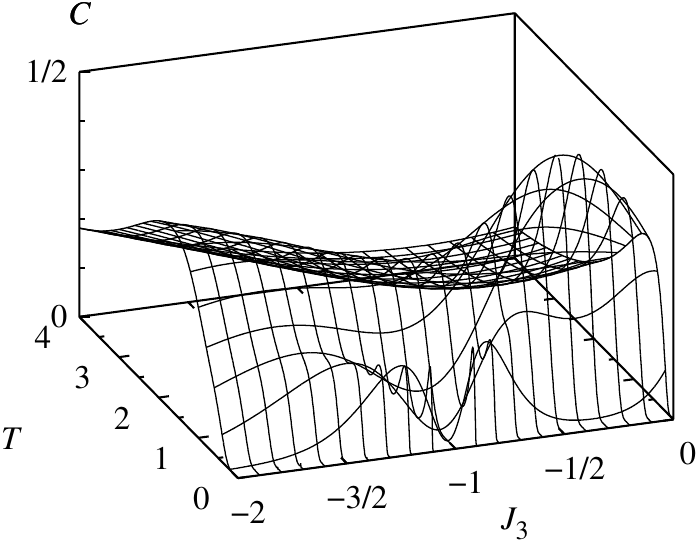}
\protect\caption{The temperature evolution of the heat capacity (\ref{eq:CV}) of the Ising chain in the antiferro-antiferro-antiferromagnetic variant of
the parameters of the exchange interactions of spins, where $J_{1}=J_{2}=-1$,
and $J_{3}<0$}
\label{fig:N3:CV:P1:mmm100} 
\end{figure}

\begin{figure}[ht]
\centering \includegraphics[scale=1.1]{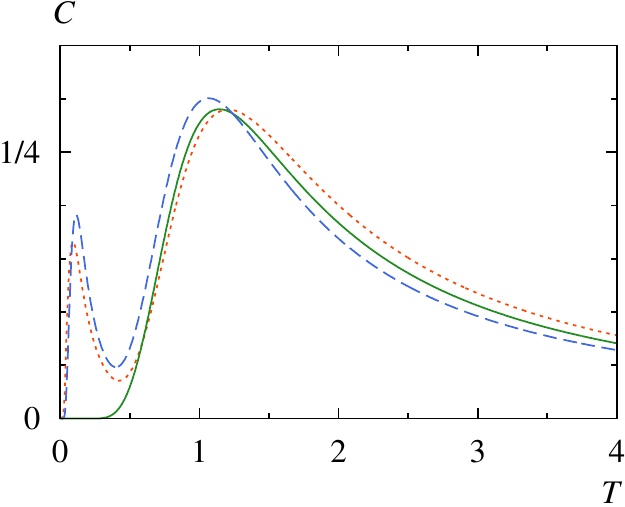}
\protect\caption{The temperature evolution of the heat capacity (\ref{eq:CV}) of the Ising chain in the vicinity of the frustration point (\ref{eq:N3:PD:P1})
of the system, where the parameters of the exchange interactions of
spins are $J_{1}=J_{2}=-1$, and the values of $J_{3}=\{-1.1,-1,-0.9\}$
correspond to the dotted, solid, and dashed lines}
\label{fig:N3:CV:P1:mmm100:} 
\end{figure}

In the considered range of model parameters, the temperature dependence
of the heat capacity of the system far from the frustration position
(\ref{eq:N3:PD:P1}) has one broad maximum, and when approaching the
frustration point, this peak splits and an additional sharp peak forms
at low temperatures, as shown in Fig.~\ref{fig:N3:CV:P1:mmm100}.
In the vicinity of the frustration regime, the sharp peak (increasing
in amplitude and decreasing in width) disappears in full at the frustration
point, where only one broad maximum remains (see Fig.~\ref{fig:N3:CV:P1:mmm100:})
\citep{Zarubin:2019:E}.

Thus, the \emph{first scenario} of the temperature evolution of the
heat capacity peaks in the frustration regime in the system is realized.

\begin{figure}[ht]
\centering \includegraphics[scale=1.1]{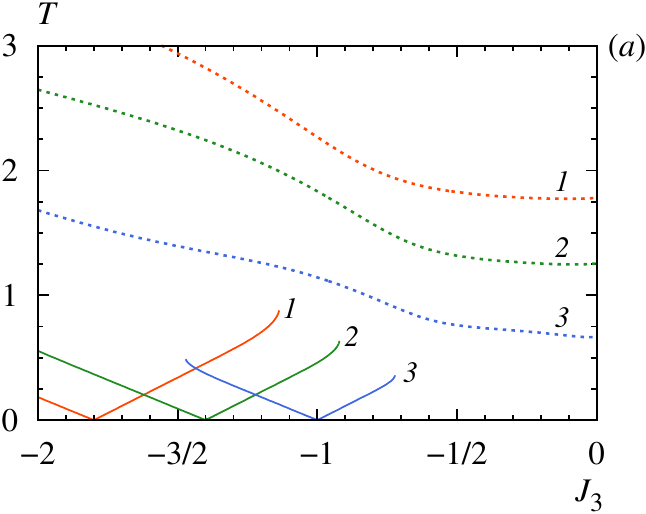}\quad{}\includegraphics[scale=1.1]{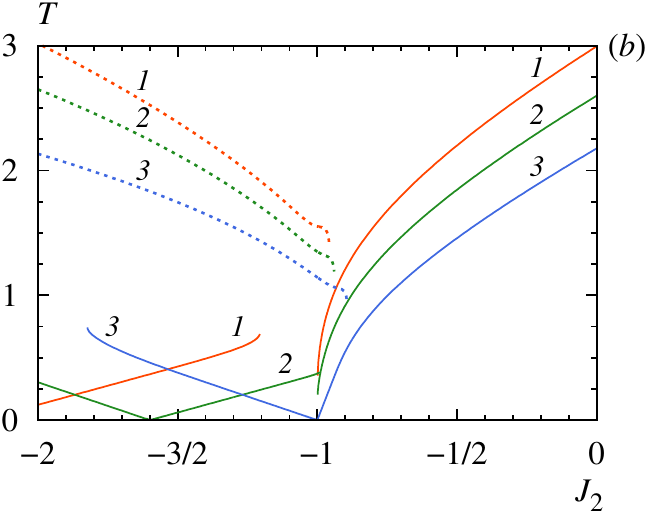}
\protect\caption{The temperature dependence of the positions of the peaks of the heat capacity (\ref{eq:CV}) of the Ising chain in the antiferro-antiferro-antiferromagnetic variant of the parameters of the exchange interactions of spins ($J_{1}=-1$, $J_{2}<0$, $J_{3}<0$) at the ratio of quantities ($J_{2}\leqslant J_{1}$ and $J_{3}\leqslant J_{1}$), where the numbering of the lines corresponds
to the sequential order of values of parameters, where $J_{2}=\{-2.2,-1.6,-1\}$
for the upper (a), and $J_{3}=\{-1.8,-1.4,-1\}$ for the lower (b)
graphs}
\label{fig:N3:CV2:L7:mmm} 
\end{figure}

\begin{figure}[ht]
\centering \includegraphics[scale=1.1]{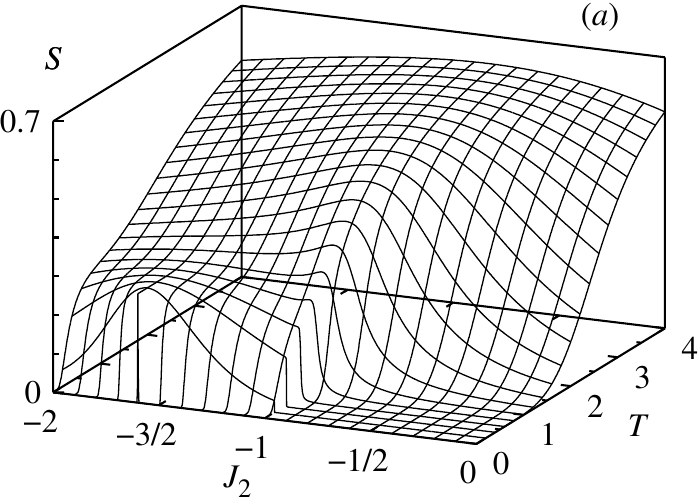}\quad{}\includegraphics[scale=1.1]{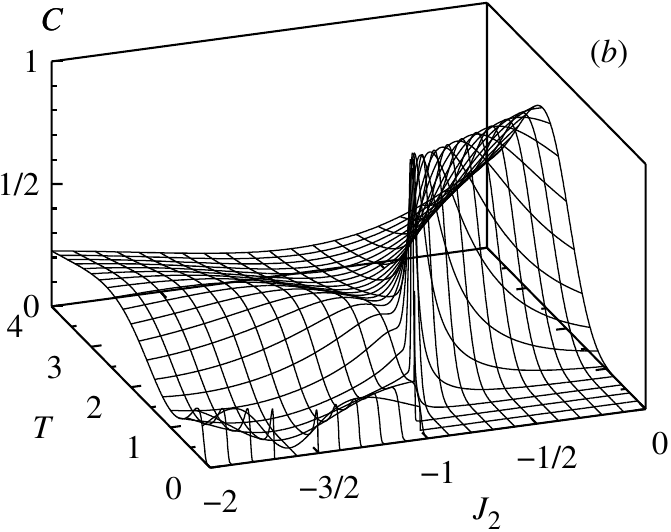}
\protect\caption{The temperature evolution of the entropy (a) and heat capacity (b)
of the Ising chain in the antiferro-antiferro-antiferromagnetic variant
of the parameters of the exchange interactions of spins, where $J_{1}=-1$,
$J_{2}<0$, and $J_{3}=-1.4$}
\label{fig:N3:TD:L7:mmm140} 
\end{figure}

The evolution of the positions of the heat capacity peaks versus temperature
in the range of model parameters (\ref{eq:N3:PD:L7}) for several
values of the exchange interaction parameters is shown in Fig.~\ref{fig:N3:CV2:L7:mmm}.
The behavior of the low-temperature peak in this figure is indicated
by a solid line, and the evolution of the peak formed at somewhat
higher temperatures is shown by a dashed line.

The most distinctly the first scenario of the temperature behavior
of the heat capacity peaks is shown in Fig.~\ref{fig:N3:CV2:L7:mmm}a.

Note that in Fig.~\ref{fig:N3:CV2:L7:mmm}b a more complicated behavior
of the heat capacity peaks was demonstrated, since in the considered
range of exchange interaction parameter values, the rearrangement
of the ordering of the spin configuration of the ground state occurs
twice (for the ratios of the model parameters (\ref{eq:N3:PD:L7})
and (\ref{eq:N3:PD:L4})). In the first case (\ref{eq:N3:PD:L7}),
the system experiences frustrations at spin ordering rearrangement,
and in the second case (\ref{eq:N3:PD:L4}) there are no frustrations
(see Fig.~\ref{fig:N3:TD:L7:mmm140}a).

The temperature evolution of entropy and heat capacity is shown in
Fig.~\ref{fig:N3:TD:L7:mmm140} for the model parameters corresponding
to line~2 in Fig.~\ref{fig:N3:CV2:L7:mmm}b. It can be seen in Fig.~\ref{fig:N3:TD:L7:mmm140}a
that the zero-temperature entropy at the boundary of spin configurations
(\ref{eq:N3:PD:L7}) at the frustration point is not zero, but at
the boundary (\ref{eq:N3:PD:L4}) at the point where frustration does
not occur, the residual entropy is zero.

Figure~\ref{fig:N3:TD:L7:mmm140}b shows that the temperature function
of the heat capacity of the system at the phase separation boundary
(\ref{eq:N3:PD:L7}), on which frustration occurs, always has one
maximum, and when deviating from these boundaries, the heat capacity
behavior noticeably changes, forming a second peak. At the boundary
(\ref{eq:N3:PD:L4}), at which frustration of the system does not
occur, the temperature evolution of the heat capacity has a more complicated
behavior, which will be discussed later.

In the case of the antiferro-antiferro-ferromagnetic variant of the
exchange interaction parameters of the model ($J_{1}<0$, $J_{2}<0$,
$J_{3}>0$) with the ratio of the quantities ($J_{2}<J_{1}/2$), the
system has frustrations at certain ratio of the model parameters (\ref{eq:N3:PD:L6})
with a non-zero value of the residual entropy equal to (\ref{eq:N3:S0:3}).
In this range of exchange interaction parameter values, the temperature
evolution of entropy and heat capacity is qualitatively similar to
the first scenario described above.

\begin{figure}[ht]
\centering \includegraphics[scale=1.1]{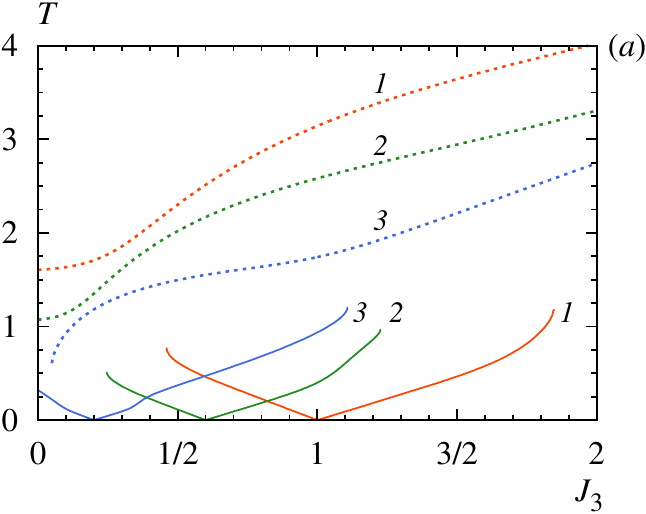}\quad{}\includegraphics[scale=1.1]{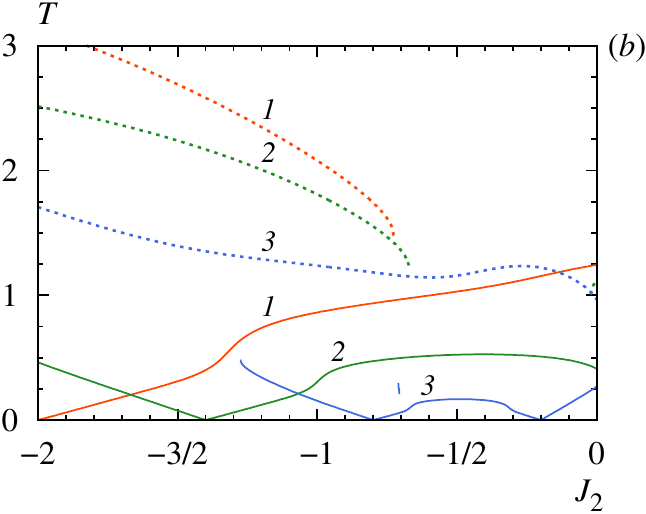}
\protect\caption{The temperature dependence of the positions of the peaks of the heat capacity (\ref{eq:CV}) of the Ising chain in the antiferro-antiferro-ferromagnetic variant of the parameters of the exchange interactions of spins ($J_{1}=-1$, $J_{2}<0$, $J_{3}>0$) for the ratio of quantities ($J_{2}<J_{1}/2$),
where the numbering of the lines corresponds to the sequential values
of parameter $J_{2}=\{-2,-1.4,-0.8\}$ for the upper (a), and $J_{3}=\{+1,+0.6,+0.2\}$ for the lower (b) graphs}
\label{fig:N3:CV2:L6:mmp} 
\end{figure}

\begin{figure}[ht]
\centering \includegraphics[scale=1.1]{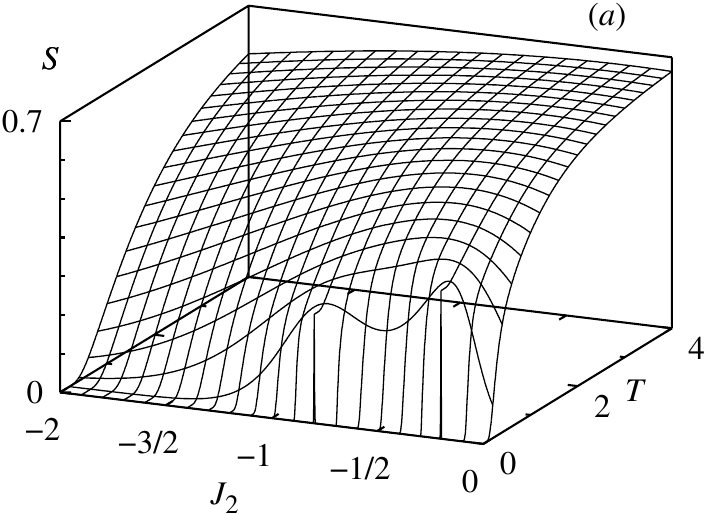}\quad{}\includegraphics[scale=1.1]{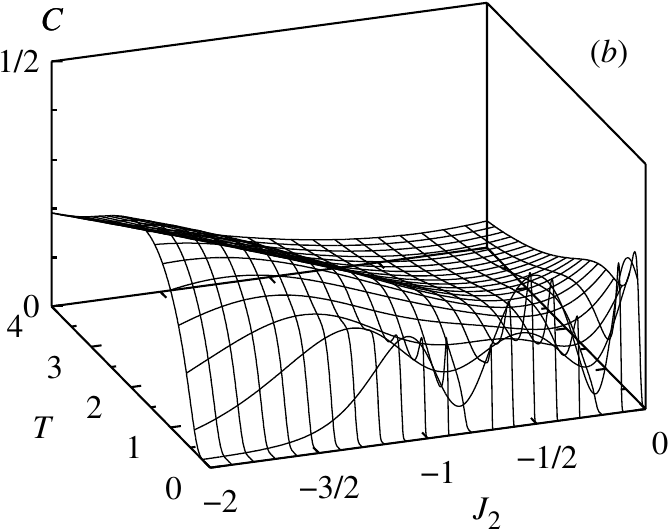}
\protect\caption{The temperature evolution of the entropy (a) and heat capacity (b)
of the Ising chain in the antiferro-antiferro-ferromagnetic variant
of the parameters of the exchange interactions of spins, where $J_{1}=-1$,
$J_{2}<0$, and $J_{3}=+0.2$}
\label{fig:N3:TD:L6:mmm020} 
\end{figure}

The temperature evolution of the positions of the heat capacity peaks
is shown in Fig.~\ref{fig:N3:CV2:L6:mmp}. As can be seen from the
heat capacity behavior in Fig.~\ref{fig:N3:CV2:L7:mmm}a and \ref{fig:N3:CV2:L6:mmp}a
the positions evolution is qualitatively similar.

It should be noted that Fig.~\ref{fig:N3:CV2:L6:mmp}b shows a complicated
evolution of the positions of the heat capacity peaks, which is associated
with the existence of two relations of the model parameters in the
range of their values at which the system experiences frustrations,
(\ref{eq:N3:PD:L6}) and (\ref{eq:N3:PD:L2}).

This situation is demonstrated by the temperature dependences of entropy
and heat capacity in Fig.~\ref{fig:N3:TD:L6:mmm020}, which correspond
to line~3 in Fig.~\ref{fig:N3:CV2:L6:mmp}b. 

The residual entropy at the boundaries of spin configurations (\ref{eq:N3:PD:L6})
and (\ref{eq:N3:PD:L2}) at the frustration points is not equal to
zero, as can be seen in Fig.~\ref{fig:N3:TD:L6:mmm020}a.

Figure~\ref{fig:N3:TD:L6:mmm020}b shows that the temperature dependence
of the heat capacity of the system at the phase separation boundaries
(\ref{eq:N3:PD:L6}) and (\ref{eq:N3:PD:L2}), on which frustrations
arise, always has only one maximum, and when deviating from these
boundaries, the heat capacity behavior changes, forming another peak
in the vicinity of the boundary.

The \emph{second scenario} of the formation of the temperature dependence
of the heat capacity of the system occurs in the antiferro-antiferro-ferromagnetic
variant of the exchange interaction parameters of the model ($J_{1}<0$,
$J_{2}<0$, $J_{3}>0$). The frustrations exist in the considered
interval of values of the variables of the model.

At first, frustrations arise at a point corresponding to the ratio
of the exchange interaction parameters of the model ($J_{2}=J_{1}/2$
and $J_{3}=0$), at which the value of residual entropy is greater
than zero and equal to the value (\ref{eq:N3:S0:1}).

Secondly, frustrations exist in the range of exchange interaction
parameters ($J_{1}/2<J_{2}\leqslant0$ and $0<J_{3}\leqslant-J_{1}/3$)
on the phase separation boundary determined by the ratio (\ref{eq:N3:PD:L2}),
while the residual entropy is equal to the value~(\ref{eq:N3:S0:2}).

In this case, in the range of model parameters ($J_{1}/2\leqslant J_{2}\leqslant0$
and $0\leqslant J_{3}\leqslant-J_{1}/3$), the temperature behavior
of the heat capacity peaks in the vicinity of the system frustration
regime is different from the first scenario described earlier. In
this case, far from the frustration regime, the temperature dependence
of the heat capacity has only one maximum, and when approaching the
frustration point, this peak splits into sharp and broad maxima, and
a broad maximum appears and forms already. Also, when approaching
the frustration point, these heat capacity maxima diverge and the
sharp peak disappears at the frustration point, and when moving away
from this point the broad maximum disappears.

\begin{figure}[ht]
\centering \includegraphics[scale=1.1]{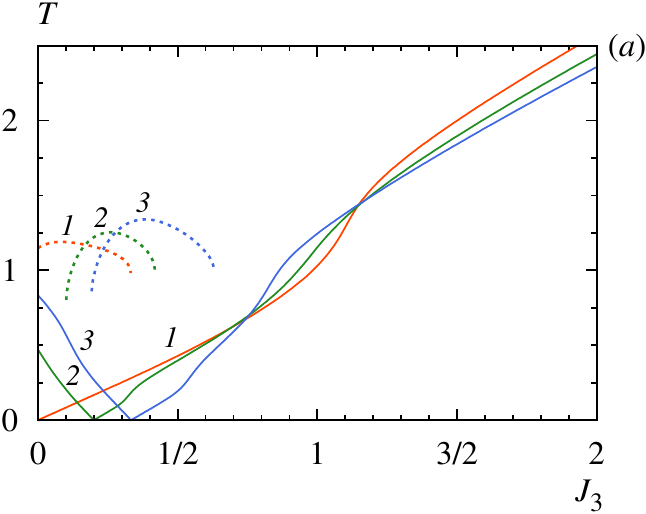}\quad{}\includegraphics[scale=1.1]{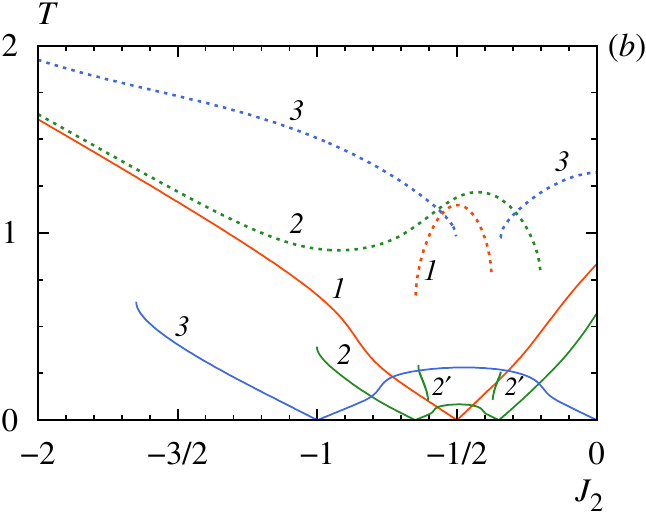}
\protect\caption{The temperature dependence of the positions of the peaks of the heat capacity (\ref{eq:CV}) of the Ising chain in the antiferro-antiferro-ferromagnetic variant of the parameters of the exchange interactions of spins ($J_{1}=-1$, $J_{2}<0$, $J_{3}>0$) for the ratio of quantities ($J_{1}/2<J_{2}\leqslant0$ and $0<J_{3}\leqslant-J_{1}/3$), where the numbering of the lines
corresponds to the sequential values of parameter $J_{2}=\{-0.5,-0.2,0\}$
for the upper (a), and $J_{3}=\{0,+0.1,+1/3\}$ for the lower (b)
graphs}
\label{fig:N3:CV2:L2m:mmp} 
\end{figure}

The second scenario of the formation and behavior of the temperature
evolution of the peaks of the heat capacity of the system considered
here is shown in Fig.~\ref{fig:N3:CV2:L2m:mmp}.

Figure~\ref{fig:N3:CV2:L2m:mmp}b shows a more complicated behavior
of the positions of the heat capacity peaks, which is associated with
the existence of two relations of the model parameters in this region
of their values, for which there are frustrations in the system, (\ref{eq:N3:PD:L6})
and (\ref{eq:N3:PD:L2}). In the considered range of model parameter
values, the temperature dependences of entropy and heat capacity were
already shown in Fig.~\ref{fig:N3:TD:L6:mmm020}, for line~2 in
Fig.~\ref{fig:N3:CV2:L2m:mmp}b.

\begin{figure}[ht]
\centering \includegraphics[scale=1.1]{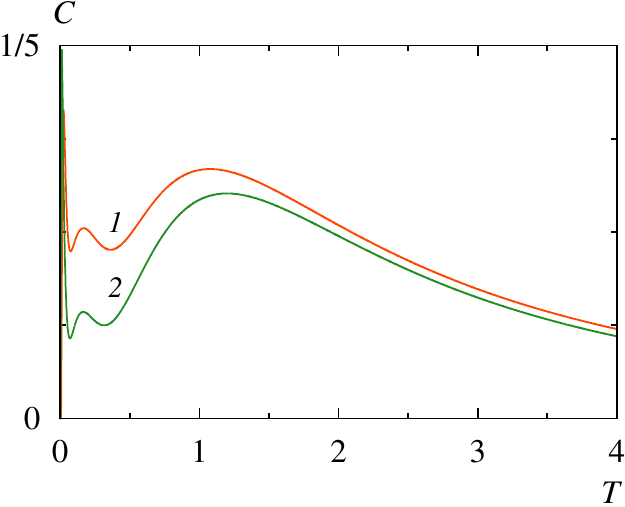}
\protect\caption{The temperature dependence of the heat capacity of the Ising chain
(\ref{eq:CV}) in the vicinity of the frustration points (\ref{eq:N3:PD:L6})
and (\ref{eq:N3:PD:L2}) of the system, where the parameters of the
exchange interactions of spins are equal to $J_{1}=-1$, $J_{3}=+0.1$,
and the values of $J_{2}=-0.612$ and $J_{2}=-0.365$ correspond to
lines 1 and 2 on the graph}
\label{fig:N3:CV:L6-L2:mmp010:3} 
\end{figure}

It should be noted here that a third peak may appear on the temperature
evolution of the heat capacity in the antiferro-antiferro-ferromagnetic
variant of the model exchange interaction parameters ($J_{1}<0$,
$J_{2}<0$, $J_{3}>0$) in the range from (\ref{eq:N3:PD:L6}) to
(\ref{eq:N3:PD:L2}), i.e. near the existence of two frustrated states
of the spin system (see lines~2' in Fig.~\ref{fig:N3:CV2:L2m:mmp}).
This small broad maximum arising between the sharp and large broad
peaks is shown in Fig.~\ref{fig:N3:CV:L6-L2:mmp010:3}.

The third scenario arises in the antiferro-ferro-ferromagnetic variant
of the exchange interaction parameters of the model ($J_{1}<0$, $J_{2}>0$,
$J_{3}>0$) for the ratios ($0<J_{2}\leqslant-J_{1}$ and $-J_{1}/3<J_{3}\leqslant-J_{1}$).
In this range of model parameters with the ratios of exchange interactions
(\ref{eq:N3:PD:L2}), the residual entropy is equal to a nonzero value
(\ref{eq:N3:S0:2}), which demonstrates the existence of frustration
in the system.

\begin{figure}[ht]
\centering \includegraphics[scale=1.1]{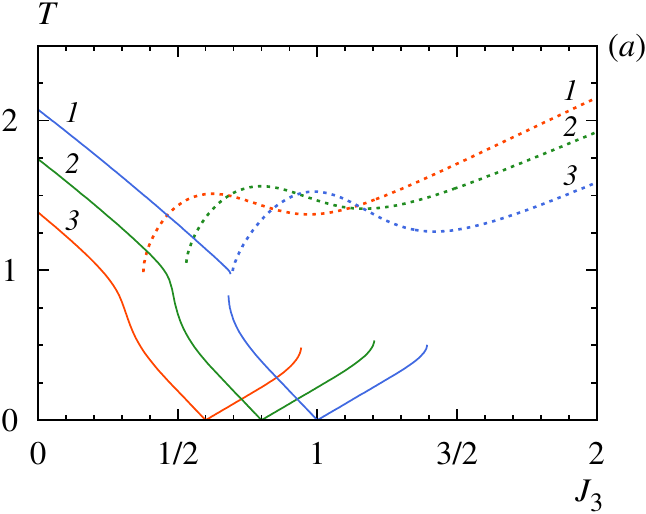}\quad{}\includegraphics[scale=1.1]{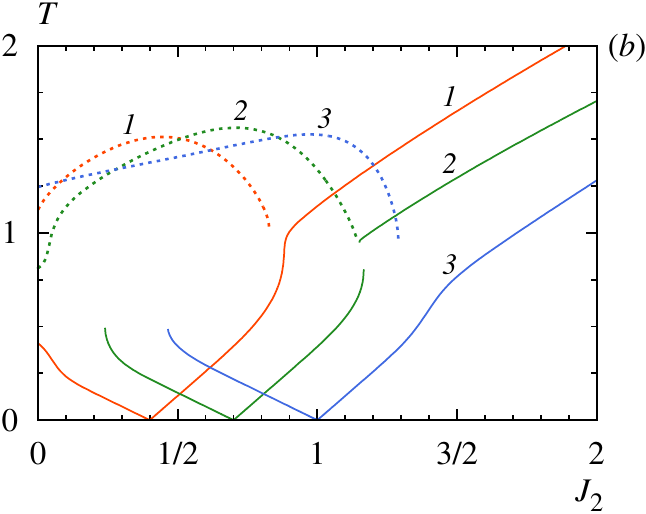}
\protect\caption{The temperature dependence of the positions of the peaks of the heat capacity (\ref{eq:CV}) of the Ising chain in the antiferro-ferro-ferromagnetic
variant of the parameters of the exchange interactions of spins ($J_{1}=-1$,
$J_{2}>0$, $J_{3}>0$) for the ratio of quantities ($0<J_{2}\leqslant-J_{1}$
and $-J_{1}/3<J_{3}\leqslant-J_{1}$), where the numbering of the
lines corresponds to the sequential values of parameter $J_{2}=\{+0.4,+0.7,+1\}$
for the upper (a), and $J_{3}=\{+0.6,+0.8,+1\}$ for the lower (b)
graphs}
\label{fig:N3:CV2:L2p:mpp} 
\end{figure}

In this case, the temperature evolution of the heat capacity peaks
demonstrates somewhat different behavior from the previously described
scenarios. When approaching the position of frustration (\ref{eq:N3:PD:L2}),
the peak splits into a broad and sharp peaks, while the sharp peak
formed at low temperatures disappears at the point of frustration.
When moving away from the frustration point towards small values of
the exchange interaction between the spins at the sites of third neighbors
($J_{3}$), the broad peak disappears (see Fig. \ref{fig:N3:CV2:L2p:mpp}a),
and when moving away from this point towards large values of the parameter
$J_{3}$, the sharp peak disappears, and only one broad maximum of
the function remains. When parameter $J_{2}$ changes (see Fig. \ref{fig:N3:CV2:L2p:mpp}b),
the temperature evolution of the heat capacity peaks is similar to
that described earlier, but with inverted behavior.

Obviously, the heat capacity of the system here demonstrates the \emph{third
scenario} of the formation of the temperature evolution of the behavior
of the maxima. This behavior is a combination of the two previous
scenarios.

This is shown that in Fig.~\ref{fig:N3:CV2:L2p:mpp} on one side
of the frustration point (\ref{eq:N3:PD:L2}) for small values of
$J_{3}$ (or large for $J_{2}$) the heat capacity behaves similarly
to the second scenario of peak development (see Fig.~\ref{fig:N3:CV2:L2m:mmp}),
and for values of $J_{3}$ greater (or less for $J_{2}$) than frustration
values (\ref{eq:N3:PD:L2}), the heat capacity behaves already as
in the case of the first scenario (see Fig.\ \ref{fig:N3:CV2:L6:mmp}).

Note that the peak splitting described above in three scenarios during
the temperature evolution of the magnetic contribution to heat capacity
is observed in real rare-earth antiferromagnets \citep{Wada:1995,Berton:1985,Avila:2004,Goruganti:2008,Chang:2010,Romero:2013,Diep:2013,Kamikawa:2015,Muller:2015}
and actinide compounds \citep{Santini:1999}, as well as a number
of organometallic coordination polymers \citep{Tran:2008,Rams:2017},
molecular \citep{Fu:2010} and quasi-one-dimensional frustrated \citep{Ahmed:2015}
magnets.

It is important to note that the magnetic phase diagram of the ground
state also contains the boundaries of the regions of spin configurations
on which frustration does not occur (dashed lines in Fig.~\ref{fig:N3:PD}).
In such cases, the entropy of the ground state of the system is zero
(\ref{eq:S0:0}), but the rearrangement of the configurations still
leads to a change in the behavior of the temperature dependence of
the heat capacity peaks of the system, which is quite different from
that described above.

\begin{figure}[ht]
\centering \includegraphics[scale=1.1]{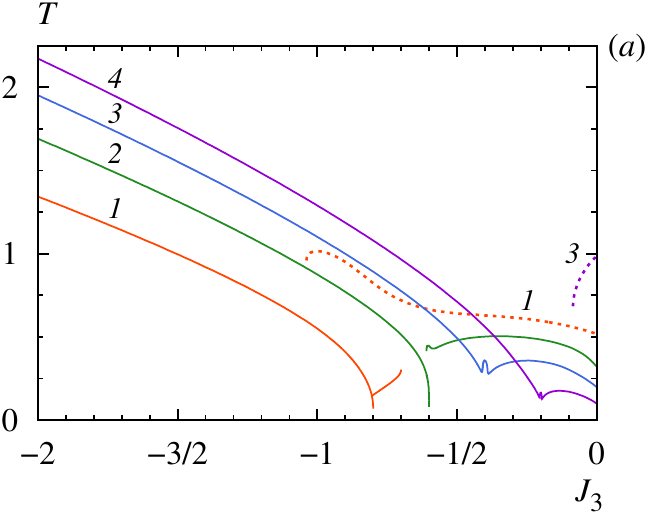}\quad{}\includegraphics[scale=1.1]{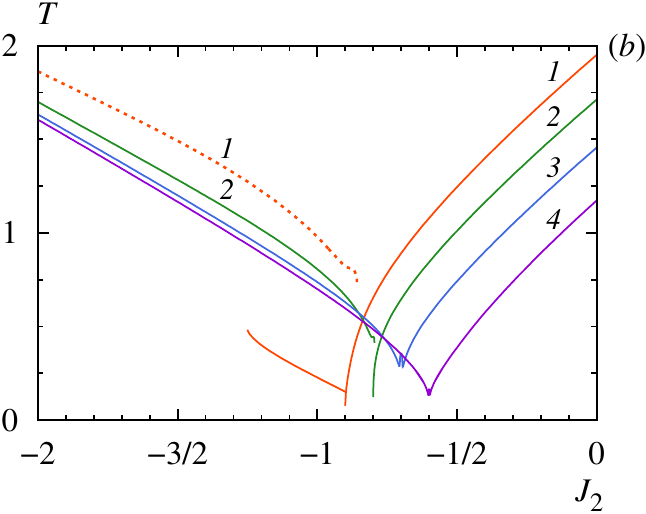}
\protect\caption{The temperature dependence of the positions of the peaks of the heat capacity (\ref{eq:CV}) of the Ising chain in the antiferro-antiferro-antiferromagnetic variant of the parameters of the exchange interactions of spins ($J_{1}=-1$, $J_{2}<0$, $J_{3}<0$) for the ratio of quantities ($J_{1}<J_{2}<J_{1}/2$ and $J_{1}<J_{3}<0$), where the numbering of the lines corresponds
to the sequential values of parameter $J_{2}=\{-0.9,-0.8,-0.7,-0.6\}$
for the upper (a), and $J_{3}=\{-0.8,-0.6,-0.4,-0.2\}$ for the lower
(b) graphs}
\label{fig:N3:CV2:L3:mmm} 
\end{figure}

These situations are observed in the magnetic phase diagram in several
cases. Firstly, in the antiferro-antiferro-anti\-ferro\-magnetic variant
of the exchange interaction parameters of the model ($J_{1}<0$, $J_{2}<0$,
$J_{3}<0$) for the ratio of quantities ($J_{1}<J_{2}<J_{1}/2$ and
$J_{1}<J_{3}<0$) at the boundary of the regions of spin configurations
(\ref{eq:N3:PD:L3}), as well as for the ratio of quantities ($J_{2}=J_{1}$
and $J_{3}<-|J_{1}|$) on the boundary (\ref{eq:N3:PD:L4}). The behavior
of the heat capacity peaks is shown in Fig.~\ref{fig:N3:CV2:L3:mmm}
and \ref{fig:N3:CV2:L7:mmm}.

\begin{figure}[ht]
\centering \includegraphics[scale=1.1]{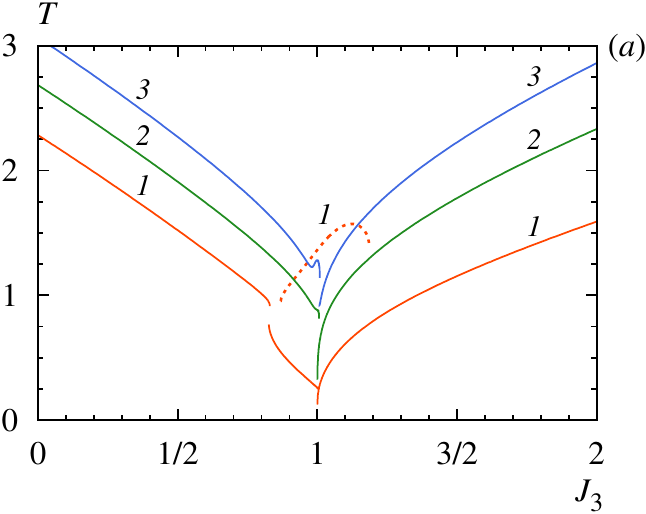}\quad{}\includegraphics[scale=1.1]{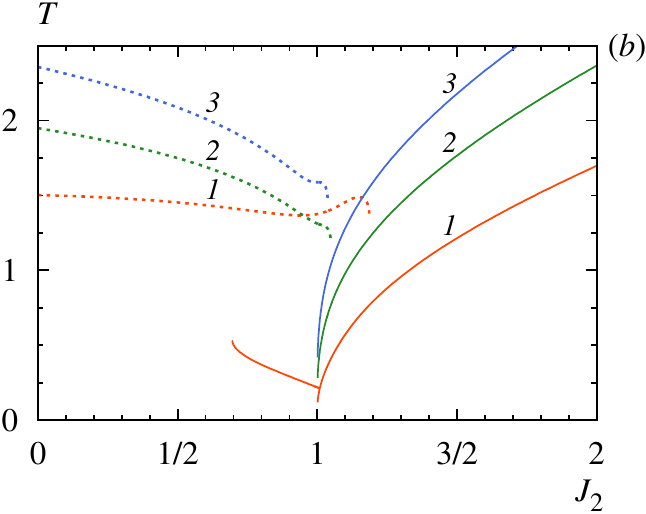}
\protect\caption{The temperature dependence of the positions of the peaks of the heat capacity (\ref{eq:CV}) of the Ising chain in the antiferro-ferro-ferromagnetic
variant of the parameters of the exchange interactions of spins ($J_{1}=-1$,
$J_{2}>0$, $J_{3}>0$) for the ratio of quantities ($J_{2}>-J_{1}$
and $J_{3}>-J_{1}$), where the numbering of the lines corresponds
to the sequential values of parameter $J_{2}=\{+1.2,+1.6,+2\}$ for
the upper (a), and $J_{3}=\{+1.2,+1.6,+2\}$ for the lower (b) graphs}
\label{fig:N3:CV2:L1-L5:mpp} 
\end{figure}

\begin{figure}[ht]
\centering \includegraphics[scale=1.1]{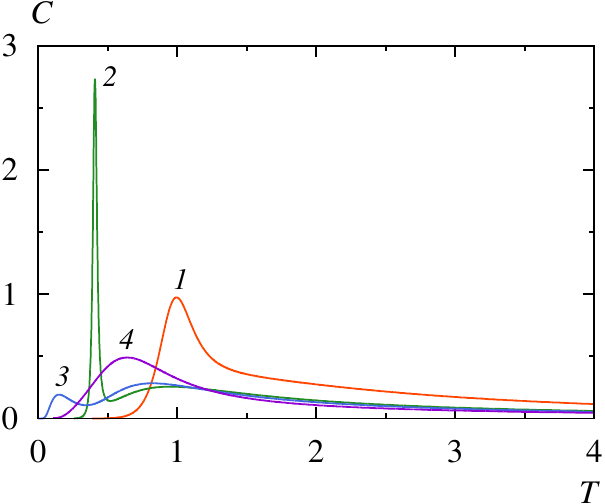}
\protect\caption{The temperature evolution of the heat capacity (\ref{eq:CV}) of the Ising chain near the magnetic phase boundary, where the ratios of
the parameters of the exchange interactions of spins are $J_{1}=-1$
and $J_{2}=-0.9$, and the line numbering corresponds to the sequential
values of the parameter $J_{3}=\{-1.5,-0.9,-0.8,-0.5\}$}
\label{fig:N3:CV:L3:mmm090:4} 
\end{figure}

Secondly, frustrations are absent in the antiferro-ferro-ferro\-magnetic
variant of the exchange interaction parameters of the model ($J_{1}<0$,
$J_{2}>0$, $J_{3}>0$) for the ratio of quantities ($J_{2}>-J_{1}$
and $J_{3}>-J_{1}$) at the boundaries of the configuration regions
of the ground state (\ref{eq:N3:PD:L1}) and (\ref{eq:N3:PD:L5}).
The behavior of the heat capacity peaks is shown in Fig.~\ref{fig:N3:CV2:L1-L5:mpp}.

\begin{figure}[ht]
\centering \includegraphics[scale=1.1]{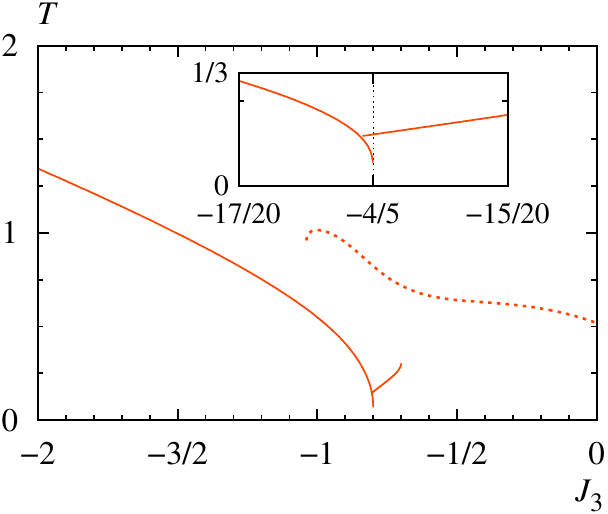}
\protect\caption{The temperature dependence of the positions of the peaks of the heat capacity (\ref{eq:CV}) of the Ising chain in the antiferro-ferro-ferromagnetic
variant of the parameters of the exchange interactions of spins ($J_{1}=-1$,
$J_{2}>0$, $J_{3}>0$) near the magnetic phase boundary $C_{\text{A}4}$--$C_{\text{A}2}$ (see Fig.~\ref{fig:N3:PD}a), where the ratio of the parameters of
the exchange interactions of spins are $J_{1}=-1$, $J_{2}=-0.9$,
and $J_{3}<0$}
\label{fig:N3:CV:L3:mmm090:m} 
\end{figure}

\begin{figure}[ht]
\centering \includegraphics[scale=1.1]{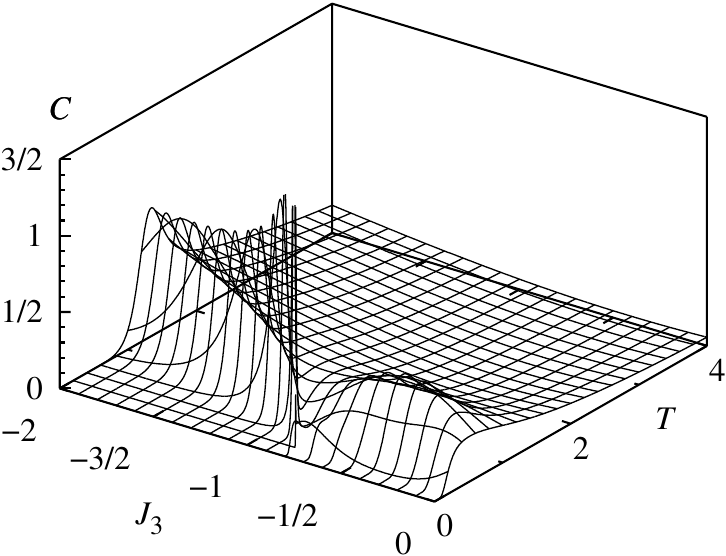}
\protect\caption{The temperature evolution of the heat capacity (\ref{eq:CV}) of the Ising chain in the antiferro-antiferro-antiferromagnetic variant of
the parameters of the exchange interactions of spins, where $J_{1}=-1$,
$J_{2}=-0.9$, and $J_{3}<0$}
\label{fig:N3:CV:L3:mmm090} 
\end{figure}

In the cases presented (with the ratio of the exchange interaction
parameters corresponding to the separation boundaries of the configurations
of the ground state considered), the position of the maximum temperature
dependence of the heat capacity formed at low temperatures does not
reach zero temperatures and does not vanished, and the heat capacity
peaks are large and small the values of the exchange interaction parameter
do not converge at one point on the phase boundary at zero temperature
(line~3 in Fig.~\ref{fig:N3:CV:L3:mmm090:4}). This behavior of
the heat capacity peaks at low temperatures demonstrates the possibility
of the formation of metastable states (see the inset in Fig.~\ref{fig:N3:CV:L3:mmm090:m}).

In the indicated ranges of the parameters of the exchange interaction
of spins at the lattice sites in the vicinity of the interphase boundary
and near the point of frustration of the system, it is possible to
split the maximum and form a two-peak structure of the temperature
dependence of the heat capacity, which differs from the heat capacity
structure near the frustration regime, as shown in Fig.~\ref{fig:N3:CV:L3:mmm090}.

This situation is possible due to the fact that the second (broad)
peak has a extension when the ratio of exchange interaction parameters
near the frustration point changes, as can be seen from Fig.~\ref{fig:N3:CV2:L7:mmm}.
In this case, the sharp peak in heat capacity can be quite high and
narrow at low temperatures and with a ratio of exchange interaction
parameters near the separation boundary of the configuration regions
(line~2 in Fig.~\ref{fig:N3:CV:L3:mmm090:4}).

In the end, it should be noted that in the antiferro-ferro-antiferromagnetic
variant of the exchange interaction parameters of the model ($J_{1}<0$,
$J_{2}>0$, $J_{3}<0$), the situation of competition between the
exchange interactions of the model does not appear and therefore changes
in the ordering of the spin configuration are not formed. The entropy
of the ground state is always zero, and the heat capacity of such
a system has only one broad peak.

\section{Conclusions}

In this work, the exact analytic expressions for the entropy and heat
capacity of the one-dimensional Ising model are obtained using the
Kramers--Wannier transfer matrix method taking into account the exchange
interactions of spins at the sites of the first, second, and third
neighbors. Based in the magnetic phase diagram of the model, the analysis
of configurational features of the ground state and frustration properties
of the system is carried out.

Criteria are formulated and relationships of model parameters are
determined for which magnetic frustrations arise in the considered
one-dimensional spin systems. It was found out that the frustrations
are caused by competition between the energies of the exchange interactions
of spins. Thus, it was shown that in the frustration regime the system
undergoes a rearrangement of the magnetic ordering structure of the
ground state, which begins to include many spin configurations comparable
to the size of the system, including the absence of translational
invariance.

The characteristic behavior of the entropy and heat capacity of the
system in the regime of frustration and near it is analyzed, a cardinal
difference in the behavior of the magnetic system in the frustration
region and outside it is shown.

It has been determined that the most important evidence of the existence
of magnetic frustrations in the system is the nonzero value of the
residual entropy in this regime, and this property does not contradict
the third law of thermodynamics. It is also shown that the residual
entropy can have the same nonzero value for entire intervals of model
parameter values.

It was found that one of the common features of frustrated systems
is the effect of peak splitting on the temperature evolution of the
magnetic contribution of the heat capacity in the immediate vicinity
of the frustration regime, which can be observed in real antiferromagnets
based on rare-earth metals and actinide compounds, as well as a number
of organometallic coordination polymers, molecular and quasi-one-dimensional
frustrated magnets.

Differences in the formation of the structure of the temperature evolution
of the heat capacity are revealed, which demonstrate three scenarios
in the behavior of the heat capacity peaks in the frustration region
of the system for various values of the ratios of the exchange interaction
parameters of the model.

The most important difference in the temperature dependence of the
heat capacity in the region of frustration and outside it lies in
the specific features of the behavior of the heat capacity peak at
low temperatures with the ratio of the model parameters corresponding
to the boundaries of the spin configurations. In the cases under consideration,
out of frustration regime the indicated heat capacity peak does not
reach zero temperatures and does not disappear at the separation boundary
of spin configurations. This peak can have a large hight, which distinguishes
it from the corresponding peaks of heat capacity in the region of
existence of system frustrations.

It is shown in the work that the heat capacity of the system near
the boundary of the phase diagram outside the frustration region demonstrates
a transition to the metastable state.

Thus, the proposed analysis scheme allows us to consider a wide range
of phenomena in one-dimensional (or quasi-one-dimensional) magnetic
systems with frustrations and to describe their relationship with
the features of thermodynamic functions. The mathematical apparatus
developed in the work makes it possible to solve similar problems
in more complicated models of statistical physics, in particular,
in multicomponent spin models with discrete symmetry and an arbitrary
spin value. 

\section*{Acknowledgment}

The research was carried out within the state assignment of Minobrnauki
of Russia (theme ``Quantum'' No.~AAAA-A18-118020190095-4), supported
in part by Ural Branch of Russian Academy of Sciences (project No.~18-2-2-11).


\begin{thebibliography}{53}%
\makeatletter
\providecommand \@ifxundefined [1]{%
 \@ifx{#1\undefined}
}%
\providecommand \@ifnum [1]{%
 \ifnum #1\expandafter \@firstoftwo
 \else \expandafter \@secondoftwo
 \fi
}%
\providecommand \@ifx [1]{%
 \ifx #1\expandafter \@firstoftwo
 \else \expandafter \@secondoftwo
 \fi
}%
\providecommand \natexlab [1]{#1}%
\providecommand \enquote  [1]{``#1''}%
\providecommand \bibnamefont  [1]{#1}%
\providecommand \bibfnamefont [1]{#1}%
\providecommand \citenamefont [1]{#1}%
\providecommand \href@noop [0]{\@secondoftwo}%
\providecommand \href [0]{\begingroup \@sanitize@url \@href}%
\providecommand \@href[1]{\@@startlink{#1}\@@href}%
\providecommand \@@href[1]{\endgroup#1\@@endlink}%
\providecommand \@sanitize@url [0]{\catcode `\\12\catcode `\$12\catcode
  `\&12\catcode `\#12\catcode `\^12\catcode `\_12\catcode `\%12\relax}%
\providecommand \@@startlink[1]{}%
\providecommand \@@endlink[0]{}%
\providecommand \url  [0]{\begingroup\@sanitize@url \@url }%
\providecommand \@url [1]{\endgroup\@href {#1}{\urlprefix }}%
\providecommand \urlprefix  [0]{URL }%
\providecommand \Eprint [0]{\href }%
\providecommand \doibase [0]{http://dx.doi.org/}%
\providecommand \selectlanguage [0]{\@gobble}%
\providecommand \bibinfo  [0]{\@secondoftwo}%
\providecommand \bibfield  [0]{\@secondoftwo}%
\providecommand \translation [1]{[#1]}%
\providecommand \BibitemOpen [0]{}%
\providecommand \bibitemStop [0]{}%
\providecommand \bibitemNoStop [0]{.\EOS\space}%
\providecommand \EOS [0]{\spacefactor3000\relax}%
\providecommand \BibitemShut  [1]{\csname bibitem#1\endcsname}%
\let\auto@bib@innerbib\@empty
\bibitem [{\citenamefont {Kassan-Ogly}\ and\ \citenamefont
  {Filippov}(2010)}]{Kassan-Ogly:2010:}%
  \BibitemOpen
  \bibfield  {author} {\bibinfo {author} {\bibfnamefont {F.~A.}\ \bibnamefont
  {Kassan-Ogly}}\ and\ \bibinfo {author} {\bibfnamefont {B.~N.}\ \bibnamefont
  {Filippov}},\ }\href {\doibase 10.3103/S1062873810100394} {\bibfield
  {journal} {\bibinfo  {journal} {Bull. Russ. Acad. Sci. Phys.}\ }\textbf
  {\bibinfo {volume} {74}},\ \bibinfo {pages} {1452} (\bibinfo {year}
  {2010})}\BibitemShut {NoStop}%
\bibitem [{\citenamefont {Normand}(2009)}]{Normand:2009}%
  \BibitemOpen
  \bibfield  {author} {\bibinfo {author} {\bibfnamefont {B.}~\bibnamefont
  {Normand}},\ }\href {\doibase 10.1080/00107510902850361} {\bibfield
  {journal} {\bibinfo  {journal} {Contemp. Phys.}\ }\textbf {\bibinfo {volume}
  {50}},\ \bibinfo {pages} {533} (\bibinfo {year} {2009})}\BibitemShut
  {NoStop}%
\bibitem [{\citenamefont {Balents}(2010)}]{Balents:2010}%
  \BibitemOpen
  \bibfield  {author} {\bibinfo {author} {\bibfnamefont {L.}~\bibnamefont
  {Balents}},\ }\href {\doibase 10.1038/nature08917} {\bibfield  {journal}
  {\bibinfo  {journal} {Nature}\ }\textbf {\bibinfo {volume} {464}},\ \bibinfo
  {pages} {199} (\bibinfo {year} {2010})}\BibitemShut {NoStop}%
\bibitem [{\citenamefont {Lacroix}\ \emph {et~al.}(2011)\citenamefont
  {Lacroix}, \citenamefont {Mendels},\ and\ \citenamefont
  {Mila}}]{Lacroix:2011}%
  \BibitemOpen
  \bibinfo {editor} {\bibfnamefont {C.}~\bibnamefont {Lacroix}}, \bibinfo
  {editor} {\bibfnamefont {P.}~\bibnamefont {Mendels}}, \ and\ \bibinfo
  {editor} {\bibfnamefont {F.}~\bibnamefont {Mila}},\ eds.,\ \href {\doibase
  10.1007/978-3-642-10589-0} {\emph {\bibinfo {title} {Introduction to
  frustrated magnetism: Materials, experiments, theory}}}\ (\bibinfo
  {publisher} {Springer},\ \bibinfo {address} {Berlin, Heidelberg},\ \bibinfo
  {year} {2011})\BibitemShut {NoStop}%
\bibitem [{\citenamefont {Sadoc}\ and\ \citenamefont
  {Mosseri}(1999)}]{Sadoc:1999}%
  \BibitemOpen
  \bibfield  {author} {\bibinfo {author} {\bibfnamefont {J.-F.}\ \bibnamefont
  {Sadoc}}\ and\ \bibinfo {author} {\bibfnamefont {R.}~\bibnamefont
  {Mosseri}},\ }\href {\doibase 10.1017/CBO9780511599934} {\emph {\bibinfo
  {title} {Geometrical frustration}}}\ (\bibinfo  {publisher} {Cambridge
  University Press},\ \bibinfo {address} {New York},\ \bibinfo {year}
  {1999})\BibitemShut {NoStop}%
\bibitem [{\citenamefont {Kudasov}\ \emph {et~al.}(2012)\citenamefont
  {Kudasov}, \citenamefont {Korshunov}, \citenamefont {Pavlov},\ and\
  \citenamefont {Maslov}}]{Kudasov:2012:}%
  \BibitemOpen
  \bibfield  {author} {\bibinfo {author} {\bibfnamefont {Y.~B.}\ \bibnamefont
  {Kudasov}}, \bibinfo {author} {\bibfnamefont {A.~S.}\ \bibnamefont
  {Korshunov}}, \bibinfo {author} {\bibfnamefont {V.~N.}\ \bibnamefont
  {Pavlov}}, \ and\ \bibinfo {author} {\bibfnamefont {D.~A.}\ \bibnamefont
  {Maslov}},\ }\href {\doibase 10.3367/UFNe.0182.201212a.1249} {\bibfield
  {journal} {\bibinfo  {journal} {Phys. Usp.}\ }\textbf {\bibinfo {volume}
  {55}},\ \bibinfo {pages} {1169} (\bibinfo {year} {2012})}\BibitemShut
  {NoStop}%
\bibitem [{\citenamefont {Diep}(2013)}]{Diep:2013}%
  \BibitemOpen
  \bibinfo {editor} {\bibfnamefont {H.~T.}\ \bibnamefont {Diep}},\ ed.,\ \href
  {\doibase 10.1142/8676} {\emph {\bibinfo {title} {Frustrated spin
  systems}}},\ \bibinfo {edition} {2nd}\ ed.\ (\bibinfo  {publisher} {World
  Scientific},\ \bibinfo {address} {New Jersey},\ \bibinfo {year}
  {2013})\BibitemShut {NoStop}%
\bibitem [{\citenamefont {Vasiliev}\ \emph {et~al.}(2018)\citenamefont
  {Vasiliev}, \citenamefont {Volkova}, \citenamefont {Zvereva},\ and\
  \citenamefont {Markina}}]{Vasiliev:2018:}%
  \BibitemOpen
  \bibfield  {author} {\bibinfo {author} {\bibfnamefont {A.~N.}\ \bibnamefont
  {Vasiliev}}, \bibinfo {author} {\bibfnamefont {O.~S.}\ \bibnamefont
  {Volkova}}, \bibinfo {author} {\bibfnamefont {E.~A.}\ \bibnamefont
  {Zvereva}}, \ and\ \bibinfo {author} {\bibfnamefont {M.~M.}\ \bibnamefont
  {Markina}},\ }\href {http://www.rfbr.ru/rffi/ru/books/o\_2052881} {\emph
  {\bibinfo {title} {Low dimensional magnetism}}}\ (\bibinfo  {publisher}
  {Fizmatlit},\ \bibinfo {address} {Moscow},\ \bibinfo {year} {2018})\ \bibinfo
  {note} {[in Russian]}\BibitemShut {NoStop}%
\bibitem [{\citenamefont {Toulouse}(1977)}]{Toulouse:1977:1}%
  \BibitemOpen
  \bibfield  {author} {\bibinfo {author} {\bibfnamefont {G.}~\bibnamefont
  {Toulouse}},\ }\href@noop {} {\bibfield  {journal} {\bibinfo  {journal}
  {Commun. Phys.}\ }\textbf {\bibinfo {volume} {2}},\ \bibinfo {pages} {115}
  (\bibinfo {year} {1977})}\BibitemShut {NoStop}%
\bibitem [{\citenamefont {Vannimenus}\ and\ \citenamefont
  {Toulouse}(1977)}]{Toulouse:1977:2}%
  \BibitemOpen
  \bibfield  {author} {\bibinfo {author} {\bibfnamefont {J.}~\bibnamefont
  {Vannimenus}}\ and\ \bibinfo {author} {\bibfnamefont {G.}~\bibnamefont
  {Toulouse}},\ }\href {\doibase 10.1088/0022-3719/10/18/008} {\bibfield
  {journal} {\bibinfo  {journal} {J. Phys. C: Solid State Phys.}\ }\textbf
  {\bibinfo {volume} {10}},\ \bibinfo {pages} {L537} (\bibinfo {year}
  {1977})}\BibitemShut {NoStop}%
\bibitem [{\citenamefont {Baxter}(1982)}]{Baxter:1982}%
  \BibitemOpen
  \bibfield  {author} {\bibinfo {author} {\bibfnamefont {R.~J.}\ \bibnamefont
  {Baxter}},\ }\href@noop {} {\emph {\bibinfo {title} {Exactly solved models in
  statistical mechanics}}}\ (\bibinfo  {publisher} {Academic Press},\ \bibinfo
  {address} {London},\ \bibinfo {year} {1982})\BibitemShut {NoStop}%
\bibitem [{\citenamefont {Ising}(1925)}]{Ising:1925}%
  \BibitemOpen
  \bibfield  {author} {\bibinfo {author} {\bibfnamefont {E.}~\bibnamefont
  {Ising}},\ }\href {\doibase 10.1007/BF02980577} {\bibfield  {journal}
  {\bibinfo  {journal} {Z. Physik}\ }\textbf {\bibinfo {volume} {31}},\
  \bibinfo {pages} {253} (\bibinfo {year} {1925})}\BibitemShut {NoStop}%
\bibitem [{\citenamefont {Brush}(1967)}]{Brush:1967}%
  \BibitemOpen
  \bibfield  {author} {\bibinfo {author} {\bibfnamefont {S.~G.}\ \bibnamefont
  {Brush}},\ }\href {\doibase 10.1103/RevModPhys.39.883} {\bibfield  {journal}
  {\bibinfo  {journal} {Rev. Mod. Phys.}\ }\textbf {\bibinfo {volume} {39}},\
  \bibinfo {pages} {883} (\bibinfo {year} {1967})}\BibitemShut {NoStop}%
\bibitem [{\citenamefont {Niss}(2005)}]{Niss:2005}%
  \BibitemOpen
  \bibfield  {author} {\bibinfo {author} {\bibfnamefont {M.}~\bibnamefont
  {Niss}},\ }\href {\doibase 10.1007/s00407-004-0088-3} {\bibfield  {journal}
  {\bibinfo  {journal} {Arch. Hist. Exact Sci.}\ }\textbf {\bibinfo {volume}
  {59}},\ \bibinfo {pages} {267} (\bibinfo {year} {2005})}\BibitemShut
  {NoStop}%
\bibitem [{\citenamefont {Newell}\ and\ \citenamefont
  {Montroll}(1953)}]{Newell:1953}%
  \BibitemOpen
  \bibfield  {author} {\bibinfo {author} {\bibfnamefont {G.~F.}\ \bibnamefont
  {Newell}}\ and\ \bibinfo {author} {\bibfnamefont {E.~W.}\ \bibnamefont
  {Montroll}},\ }\href {\doibase 10.1103/RevModPhys.25.353} {\bibfield
  {journal} {\bibinfo  {journal} {Rev. Mod. Phys.}\ }\textbf {\bibinfo {volume}
  {25}},\ \bibinfo {pages} {353} (\bibinfo {year} {1953})}\BibitemShut
  {NoStop}%
\bibitem [{\citenamefont {Kassan-Ogly}(2001)}]{Kassan-Ogly:2001}%
  \BibitemOpen
  \bibfield  {author} {\bibinfo {author} {\bibfnamefont {F.~A.}\ \bibnamefont
  {Kassan-Ogly}},\ }\href {\doibase 10.1080/01411590108227581} {\bibfield
  {journal} {\bibinfo  {journal} {Phase Transitions}\ }\textbf {\bibinfo
  {volume} {74}},\ \bibinfo {pages} {353} (\bibinfo {year} {2001})}\BibitemShut
  {NoStop}%
\bibitem [{\citenamefont {Kassan-Ogly}\ \emph {et~al.}(1989)\citenamefont
  {Kassan-Ogly}, \citenamefont {Kormiltsev}, \citenamefont {Naish},\ and\
  \citenamefont {Sagaradze}}]{Kassan-Ogly:1989:}%
  \BibitemOpen
  \bibfield  {author} {\bibinfo {author} {\bibfnamefont {F.~A.}\ \bibnamefont
  {Kassan-Ogly}}, \bibinfo {author} {\bibfnamefont {E.~V.}\ \bibnamefont
  {Kormiltsev}}, \bibinfo {author} {\bibfnamefont {V.~E.}\ \bibnamefont
  {Naish}}, \ and\ \bibinfo {author} {\bibfnamefont {I.~V.}\ \bibnamefont
  {Sagaradze}},\ }\href@noop {} {\bibfield  {journal} {\bibinfo  {journal}
  {Sov. Phys. Solid State}\ }\textbf {\bibinfo {volume} {31}},\ \bibinfo
  {pages} {43} (\bibinfo {year} {1989})}\BibitemShut {NoStop}%
\bibitem [{\citenamefont {Zarubin}\ \emph {et~al.}(2016)\citenamefont
  {Zarubin}, \citenamefont {Kassan-Ogly},\ and\ \citenamefont
  {Proshkin}}]{Zarubin:2016}%
  \BibitemOpen
  \bibfield  {author} {\bibinfo {author} {\bibfnamefont {A.~V.}\ \bibnamefont
  {Zarubin}}, \bibinfo {author} {\bibfnamefont {F.~A.}\ \bibnamefont
  {Kassan-Ogly}}, \ and\ \bibinfo {author} {\bibfnamefont {A.~I.}\ \bibnamefont
  {Proshkin}},\ }\href {\doibase 10.4028/www.scientific.net/MSF.845.122}
  {\bibfield  {journal} {\bibinfo  {journal} {Mater. Sci. Forum}\ }\textbf
  {\bibinfo {volume} {845}},\ \bibinfo {pages} {122} (\bibinfo {year}
  {2016})}\BibitemShut {NoStop}%
\bibitem [{\citenamefont {Kramers}\ and\ \citenamefont
  {Wannier}(1941)}]{Kramers:1941:1}%
  \BibitemOpen
  \bibfield  {author} {\bibinfo {author} {\bibfnamefont {H.~A.}\ \bibnamefont
  {Kramers}}\ and\ \bibinfo {author} {\bibfnamefont {G.~H.}\ \bibnamefont
  {Wannier}},\ }\href {\doibase 10.1103/PhysRev.60.252} {\bibfield  {journal}
  {\bibinfo  {journal} {Phys. Rev.}\ }\textbf {\bibinfo {volume} {60}},\
  \bibinfo {pages} {252} (\bibinfo {year} {1941})}\BibitemShut {NoStop}%
\bibitem [{\citenamefont {Oguchi}(1965)}]{Oguchi:1965}%
  \BibitemOpen
  \bibfield  {author} {\bibinfo {author} {\bibfnamefont {T.}~\bibnamefont
  {Oguchi}},\ }\href {\doibase 10.1143/JPSJ.20.2236} {\bibfield  {journal}
  {\bibinfo  {journal} {J.~Phys. Soc. Jpn.}\ }\textbf {\bibinfo {volume}
  {20}},\ \bibinfo {pages} {2236} (\bibinfo {year} {1965})}\BibitemShut
  {NoStop}%
\bibitem [{\citenamefont {Zarubin}\ \emph
  {et~al.}(2019{\natexlab{a}})\citenamefont {Zarubin}, \citenamefont
  {Kassan-Ogly}, \citenamefont {Proshkin},\ and\ \citenamefont
  {Shestakov}}]{Zarubin:2019:}%
  \BibitemOpen
  \bibfield  {author} {\bibinfo {author} {\bibfnamefont {A.~V.}\ \bibnamefont
  {Zarubin}}, \bibinfo {author} {\bibfnamefont {F.~A.}\ \bibnamefont
  {Kassan-Ogly}}, \bibinfo {author} {\bibfnamefont {A.~I.}\ \bibnamefont
  {Proshkin}}, \ and\ \bibinfo {author} {\bibfnamefont {A.~E.}\ \bibnamefont
  {Shestakov}},\ }\href {\doibase 10.1134/S106377611904006X} {\bibfield
  {journal} {\bibinfo  {journal} {J. Exp. Theor. Phys.}\ }\textbf {\bibinfo
  {volume} {128}},\ \bibinfo {pages} {778} (\bibinfo {year}
  {2019}{\natexlab{a}})}\BibitemShut {NoStop}%
\bibitem [{\citenamefont {Domb}(1960)}]{Domb:1960}%
  \BibitemOpen
  \bibfield  {author} {\bibinfo {author} {\bibfnamefont {C.}~\bibnamefont
  {Domb}},\ }\href {\doibase 10.1080/00018736000101189} {\bibfield  {journal}
  {\bibinfo  {journal} {Adv. Phys.}\ }\textbf {\bibinfo {volume} {9}},\
  \bibinfo {pages} {149} (\bibinfo {year} {1960})}\BibitemShut {NoStop}%
\bibitem [{\citenamefont {Horn}\ and\ \citenamefont
  {Johnson}(2013)}]{Horn:2013}%
  \BibitemOpen
  \bibfield  {author} {\bibinfo {author} {\bibfnamefont {R.~A.}\ \bibnamefont
  {Horn}}\ and\ \bibinfo {author} {\bibfnamefont {C.~R.}\ \bibnamefont
  {Johnson}},\ }\href {http://www.cambridge.org/9780521548236} {\emph {\bibinfo
  {title} {Matrix analysis}}},\ \bibinfo {edition} {2nd}\ ed.\ (\bibinfo
  {publisher} {Cambridge University Press},\ \bibinfo {address} {Cambridge},\
  \bibinfo {year} {2013})\BibitemShut {NoStop}%
\bibitem [{\citenamefont {Nolting}\ and\ \citenamefont
  {Ramakanth}(2009)}]{Nolting:2009}%
  \BibitemOpen
  \bibfield  {author} {\bibinfo {author} {\bibfnamefont {W.}~\bibnamefont
  {Nolting}}\ and\ \bibinfo {author} {\bibfnamefont {A.}~\bibnamefont
  {Ramakanth}},\ }\href {\doibase 10.1007/978-3-540-85416-6} {\emph {\bibinfo
  {title} {Quantum theory of magnetism}}}\ (\bibinfo  {publisher} {Springer},\
  \bibinfo {address} {Berlin, Heidelberg},\ \bibinfo {year} {2009})\BibitemShut
  {NoStop}%
\bibitem [{\citenamefont {Mussardo}(2010)}]{Mussardo:2010}%
  \BibitemOpen
  \bibfield  {author} {\bibinfo {author} {\bibfnamefont {G.}~\bibnamefont
  {Mussardo}},\ }\href
  {https://global.oup.com/academic/product/statistical-field-theory-9780199547586}
  {\emph {\bibinfo {title} {Statistical field theory: An introduction to
  exactly solved models in statistical physics}}}\ (\bibinfo  {publisher}
  {Oxford University Press},\ \bibinfo {address} {Oxford, New York},\ \bibinfo
  {year} {2010})\BibitemShut {NoStop}%
\bibitem [{\citenamefont {Gould}\ and\ \citenamefont
  {Tobochnik}(2010)}]{Gould:2010}%
  \BibitemOpen
  \bibfield  {author} {\bibinfo {author} {\bibfnamefont {H.}~\bibnamefont
  {Gould}}\ and\ \bibinfo {author} {\bibfnamefont {J.}~\bibnamefont
  {Tobochnik}},\ }\href {\doibase 10.2307/j.ctvcmxp2z} {\emph {\bibinfo {title}
  {Statistical and thermal physics: With computer applications}}}\ (\bibinfo
  {publisher} {Princeton University Press},\ \bibinfo {address} {Princeton},\
  \bibinfo {year} {2010})\BibitemShut {NoStop}%
\bibitem [{\citenamefont {Mart\'{\i}nez-Garcilazo}\ \emph
  {et~al.}(2009)\citenamefont {Mart\'{\i}nez-Garcilazo}, \citenamefont
  {M\'{a}rquez-Islas},\ and\ \citenamefont
  {Ram\'{\i}rez-Romero}}]{Martinez-Garcilazo:2009}%
  \BibitemOpen
  \bibfield  {author} {\bibinfo {author} {\bibfnamefont {J.~P.}\ \bibnamefont
  {Mart\'{\i}nez-Garcilazo}}, \bibinfo {author} {\bibfnamefont
  {R.}~\bibnamefont {M\'{a}rquez-Islas}}, \ and\ \bibinfo {author}
  {\bibfnamefont {C.}~\bibnamefont {Ram\'{\i}rez-Romero}},\ }\href
  {http://www.scielo.org.mx/scielo.php?script=sci\_arttext\&pid=S1870-35422009000200001}
  {\bibfield  {journal} {\bibinfo  {journal} {Rev. mex. f\'{\i}s.~E}\ }\textbf
  {\bibinfo {volume} {55}},\ \bibinfo {pages} {136} (\bibinfo {year}
  {2009})}\BibitemShut {NoStop}%
\bibitem [{\citenamefont {Stephenson}(1970)}]{Stephenson:1970}%
  \BibitemOpen
  \bibfield  {author} {\bibinfo {author} {\bibfnamefont {J.}~\bibnamefont
  {Stephenson}},\ }\href {\doibase 10.1139/p70-217} {\bibfield  {journal}
  {\bibinfo  {journal} {Can. J. Phys.}\ }\textbf {\bibinfo {volume} {48}},\
  \bibinfo {pages} {1724} (\bibinfo {year} {1970})}\BibitemShut {NoStop}%
\bibitem [{\citenamefont {Fisher}\ and\ \citenamefont
  {Selke}(1980)}]{Fisher:1980}%
  \BibitemOpen
  \bibfield  {author} {\bibinfo {author} {\bibfnamefont {M.~E.}\ \bibnamefont
  {Fisher}}\ and\ \bibinfo {author} {\bibfnamefont {W.}~\bibnamefont {Selke}},\
  }\href {\doibase 10.1103/PhysRevLett.44.1502} {\bibfield  {journal} {\bibinfo
   {journal} {Phys. Rev. Lett.}\ }\textbf {\bibinfo {volume} {44}},\ \bibinfo
  {pages} {1502} (\bibinfo {year} {1980})}\BibitemShut {NoStop}%
\bibitem [{\citenamefont {Fisher}\ and\ \citenamefont
  {Selke}(1981)}]{Fisher:1981}%
  \BibitemOpen
  \bibfield  {author} {\bibinfo {author} {\bibfnamefont {M.~E.}\ \bibnamefont
  {Fisher}}\ and\ \bibinfo {author} {\bibfnamefont {W.}~\bibnamefont {Selke}},\
  }\href {\doibase 10.1098/rsta.1981.0156} {\bibfield  {journal} {\bibinfo
  {journal} {Phil. Trans. R. Soc. A}\ }\textbf {\bibinfo {volume} {302}},\
  \bibinfo {pages} {1} (\bibinfo {year} {1981})}\BibitemShut {NoStop}%
\bibitem [{\citenamefont {Selke}\ \emph {et~al.}(1985)\citenamefont {Selke},
  \citenamefont {Barreto},\ and\ \citenamefont {Yeomans}}]{Selke:1985}%
  \BibitemOpen
  \bibfield  {author} {\bibinfo {author} {\bibfnamefont {W.}~\bibnamefont
  {Selke}}, \bibinfo {author} {\bibfnamefont {M.}~\bibnamefont {Barreto}}, \
  and\ \bibinfo {author} {\bibfnamefont {J.}~\bibnamefont {Yeomans}},\ }\href
  {\doibase 10.1088/0022-3719/18/14/007} {\bibfield  {journal} {\bibinfo
  {journal} {J. Phys. C: Solid State Phys.}\ }\textbf {\bibinfo {volume}
  {18}},\ \bibinfo {pages} {L393} (\bibinfo {year} {1985})}\BibitemShut
  {NoStop}%
\bibitem [{\citenamefont {Selke}(1988)}]{Selke:1988}%
  \BibitemOpen
  \bibfield  {author} {\bibinfo {author} {\bibfnamefont {W.}~\bibnamefont
  {Selke}},\ }\href {\doibase 10.1016/0370-1573(88)90140-8} {\bibfield
  {journal} {\bibinfo  {journal} {Phys. Reports}\ }\textbf {\bibinfo {volume}
  {170}},\ \bibinfo {pages} {213} (\bibinfo {year} {1988})}\BibitemShut
  {NoStop}%
\bibitem [{\citenamefont {Yeomans}(1988)}]{Yeomans:1988}%
  \BibitemOpen
  \bibfield  {author} {\bibinfo {author} {\bibfnamefont {J.}~\bibnamefont
  {Yeomans}},\ }\enquote {\bibinfo {title} {The theory and application of axial
  {Ising} models},}\ in\ \href {\doibase 10.1016/S0081-1947(08)60379-3} {\emph
  {\bibinfo {booktitle} {Solid state physics}}},\ Vol.~\bibinfo {volume} {41},\
  \bibinfo {editor} {edited by\ \bibinfo {editor} {\bibfnamefont
  {H.}~\bibnamefont {Ehrenreich}}\ and\ \bibinfo {editor} {\bibfnamefont
  {D.}~\bibnamefont {Turnbull}}}\ (\bibinfo  {publisher} {Academic Press},\
  \bibinfo {address} {San Diego},\ \bibinfo {year} {1988})\ pp.\ \bibinfo
  {pages} {151--200}\BibitemShut {NoStop}%
\bibitem [{\citenamefont {Price}(1983)}]{Price:1983}%
  \BibitemOpen
  \bibfield  {author} {\bibinfo {author} {\bibfnamefont {G.~D.}\ \bibnamefont
  {Price}},\ }\href {\doibase 10.1007/BF00309588} {\bibfield  {journal}
  {\bibinfo  {journal} {Phys. Chem. Miner.}\ }\textbf {\bibinfo {volume}
  {10}},\ \bibinfo {pages} {77} (\bibinfo {year} {1983})}\BibitemShut {NoStop}%
\bibitem [{\citenamefont {Barreto}\ and\ \citenamefont
  {Yeomans}(1985)}]{Barreto:1985}%
  \BibitemOpen
  \bibfield  {author} {\bibinfo {author} {\bibfnamefont {M.}~\bibnamefont
  {Barreto}}\ and\ \bibinfo {author} {\bibfnamefont {J.}~\bibnamefont
  {Yeomans}},\ }\href {\doibase 10.1016/0378-4371(85)90157-8} {\bibfield
  {journal} {\bibinfo  {journal} {Physica A}\ }\textbf {\bibinfo {volume}
  {134}},\ \bibinfo {pages} {84} (\bibinfo {year} {1985})}\BibitemShut
  {NoStop}%
\bibitem [{\citenamefont {Pokrovskii}\ and\ \citenamefont
  {Uimin}(1982)}]{Pokrovskii:1982:}%
  \BibitemOpen
  \bibfield  {author} {\bibinfo {author} {\bibfnamefont {V.~L.}\ \bibnamefont
  {Pokrovskii}}\ and\ \bibinfo {author} {\bibfnamefont {G.~V.}\ \bibnamefont
  {Uimin}},\ }\href {http://www.jetp.ac.ru/cgi-bin/r/index/e/55/5/p950?a=list}
  {\bibfield  {journal} {\bibinfo  {journal} {Sov. Phys. JETP}\ }\textbf
  {\bibinfo {volume} {55}},\ \bibinfo {pages} {950} (\bibinfo {year}
  {1982})}\BibitemShut {NoStop}%
\bibitem [{\citenamefont {Yeomans}(1987)}]{Yeomans:1987}%
  \BibitemOpen
  \bibfield  {author} {\bibinfo {author} {\bibfnamefont {J.}~\bibnamefont
  {Yeomans}},\ }\enquote {\bibinfo {title} {The application of axial ising
  models to the description of modulated order},}\ in\ \href {\doibase
  10.1007/978-1-4757-0184-5} {\emph {\bibinfo {booktitle} {Incommensurate
  crystals, liquid crystals, and quasi-crystals}}},\ \bibinfo {editor} {edited
  by\ \bibinfo {editor} {\bibfnamefont {J.~F.}\ \bibnamefont {Scott}}\ and\
  \bibinfo {editor} {\bibfnamefont {N.~A.}\ \bibnamefont {Clark}}}\ (\bibinfo
  {publisher} {Plenum Press},\ \bibinfo {address} {New York},\ \bibinfo {year}
  {1987})\BibitemShut {NoStop}%
\bibitem [{\citenamefont {Sommerfeld}(1956)}]{Sommerfeld:1956}%
  \BibitemOpen
  \bibfield  {author} {\bibinfo {author} {\bibfnamefont {A.}~\bibnamefont
  {Sommerfeld}},\ }\href@noop {} {\emph {\bibinfo {title} {Thermodynamics and
  statistical mechanics}}}\ (\bibinfo  {publisher} {Academic Press},\ \bibinfo
  {address} {New York},\ \bibinfo {year} {1956})\BibitemShut {NoStop}%
\bibitem [{\citenamefont {Nolting}(2018)}]{Nolting8:2018}%
  \BibitemOpen
  \bibfield  {author} {\bibinfo {author} {\bibfnamefont {W.}~\bibnamefont
  {Nolting}},\ }\href {\doibase 10.1007/978-3-319-73827-7} {\emph {\bibinfo
  {title} {Theoretical physics 8: Statistical physics}}}\ (\bibinfo
  {publisher} {Springer},\ \bibinfo {address} {Cham},\ \bibinfo {year}
  {2018})\BibitemShut {NoStop}%
\bibitem [{\citenamefont {Zarubin}\ \emph
  {et~al.}(2019{\natexlab{b}})\citenamefont {Zarubin}, \citenamefont
  {Kassan-Ogly},\ and\ \citenamefont {Proshkin}}]{Zarubin:2019:E}%
  \BibitemOpen
  \bibfield  {author} {\bibinfo {author} {\bibfnamefont {A.~V.}\ \bibnamefont
  {Zarubin}}, \bibinfo {author} {\bibfnamefont {F.~A.}\ \bibnamefont
  {Kassan-Ogly}}, \ and\ \bibinfo {author} {\bibfnamefont {A.~I.}\ \bibnamefont
  {Proshkin}},\ }\href {\doibase 10.1088/1742-6596/1389/1/012009} {\bibfield
  {journal} {\bibinfo  {journal} {J. Phys.: Conf. Ser.}\ }\textbf {\bibinfo
  {volume} {1389}},\ \bibinfo {pages} {012009} (\bibinfo {year}
  {2019}{\natexlab{b}})}\BibitemShut {NoStop}%
\bibitem [{\citenamefont {Wada}\ \emph {et~al.}(1995)\citenamefont {Wada},
  \citenamefont {Imai},\ and\ \citenamefont {Shiga}}]{Wada:1995}%
  \BibitemOpen
  \bibfield  {author} {\bibinfo {author} {\bibfnamefont {H.}~\bibnamefont
  {Wada}}, \bibinfo {author} {\bibfnamefont {H.}~\bibnamefont {Imai}}, \ and\
  \bibinfo {author} {\bibfnamefont {M.}~\bibnamefont {Shiga}},\ }\href
  {\doibase 10.1016/0925-8388(94)01401-9} {\bibfield  {journal} {\bibinfo
  {journal} {J. Alloys Compd.}\ }\textbf {\bibinfo {volume} {218}},\ \bibinfo
  {pages} {73} (\bibinfo {year} {1995})}\BibitemShut {NoStop}%
\bibitem [{\citenamefont {Berton}\ \emph {et~al.}(1985)\citenamefont {Berton},
  \citenamefont {Chaussy}, \citenamefont {Flouquet}, \citenamefont {Odin},
  \citenamefont {Peyrard},\ and\ \citenamefont {Holtzberg}}]{Berton:1985}%
  \BibitemOpen
  \bibfield  {author} {\bibinfo {author} {\bibfnamefont {A.}~\bibnamefont
  {Berton}}, \bibinfo {author} {\bibfnamefont {J.}~\bibnamefont {Chaussy}},
  \bibinfo {author} {\bibfnamefont {J.}~\bibnamefont {Flouquet}}, \bibinfo
  {author} {\bibfnamefont {J.}~\bibnamefont {Odin}}, \bibinfo {author}
  {\bibfnamefont {J.}~\bibnamefont {Peyrard}}, \ and\ \bibinfo {author}
  {\bibfnamefont {F.}~\bibnamefont {Holtzberg}},\ }\href {\doibase
  10.1103/PhysRevB.31.4313} {\bibfield  {journal} {\bibinfo  {journal} {Phys.
  Rev. B}\ }\textbf {\bibinfo {volume} {31}},\ \bibinfo {pages} {4313}
  (\bibinfo {year} {1985})}\BibitemShut {NoStop}%
\bibitem [{\citenamefont {Avila}\ \emph {et~al.}(2004)\citenamefont {Avila},
  \citenamefont {Bud'ko},\ and\ \citenamefont {Canfield}}]{Avila:2004}%
  \BibitemOpen
  \bibfield  {author} {\bibinfo {author} {\bibfnamefont {M.~A.}\ \bibnamefont
  {Avila}}, \bibinfo {author} {\bibfnamefont {S.~L.}\ \bibnamefont {Bud'ko}}, \
  and\ \bibinfo {author} {\bibfnamefont {P.~C.}\ \bibnamefont {Canfield}},\
  }\href {\doibase 10.1016/S0304-8853(03)00672-3} {\bibfield  {journal}
  {\bibinfo  {journal} {J. Magn. Magn. Mater.}\ }\textbf {\bibinfo {volume}
  {270}},\ \bibinfo {pages} {51} (\bibinfo {year} {2004})}\BibitemShut
  {NoStop}%
\bibitem [{\citenamefont {Goruganti}\ \emph {et~al.}(2008)\citenamefont
  {Goruganti}, \citenamefont {Rathnayaka}, \citenamefont {Ross}, \citenamefont
  {\"{O}ner}, \citenamefont {Lue},\ and\ \citenamefont {Kuo}}]{Goruganti:2008}%
  \BibitemOpen
  \bibfield  {author} {\bibinfo {author} {\bibfnamefont {V.}~\bibnamefont
  {Goruganti}}, \bibinfo {author} {\bibfnamefont {K.~D.~D.}\ \bibnamefont
  {Rathnayaka}}, \bibinfo {author} {\bibfnamefont {J.~H.}\ \bibnamefont
  {Ross}}, \bibinfo {author} {\bibfnamefont {Y.}~\bibnamefont {\"{O}ner}},
  \bibinfo {author} {\bibfnamefont {C.~S.}\ \bibnamefont {Lue}}, \ and\
  \bibinfo {author} {\bibfnamefont {Y.~K.}\ \bibnamefont {Kuo}},\ }\href
  {\doibase 10.1063/1.2904856} {\bibfield  {journal} {\bibinfo  {journal} {J.
  Appl. Phys.}\ }\textbf {\bibinfo {volume} {103}},\ \bibinfo {pages} {073919}
  (\bibinfo {year} {2008})}\BibitemShut {NoStop}%
\bibitem [{\citenamefont {Chang}\ \emph {et~al.}(2010)\citenamefont {Chang},
  \citenamefont {Prager}, \citenamefont {Per{\ss}on}, \citenamefont {Walter},
  \citenamefont {Jansen}, \citenamefont {Chen},\ and\ \citenamefont
  {Gardner}}]{Chang:2010}%
  \BibitemOpen
  \bibfield  {author} {\bibinfo {author} {\bibfnamefont {L.~J.}\ \bibnamefont
  {Chang}}, \bibinfo {author} {\bibfnamefont {M.}~\bibnamefont {Prager}},
  \bibinfo {author} {\bibfnamefont {J.}~\bibnamefont {Per{\ss}on}}, \bibinfo
  {author} {\bibfnamefont {J.}~\bibnamefont {Walter}}, \bibinfo {author}
  {\bibfnamefont {E.}~\bibnamefont {Jansen}}, \bibinfo {author} {\bibfnamefont
  {Y.~Y.}\ \bibnamefont {Chen}}, \ and\ \bibinfo {author} {\bibfnamefont
  {J.~S.}\ \bibnamefont {Gardner}},\ }\href {\doibase
  10.1088/0953-8984/22/7/076003} {\bibfield  {journal} {\bibinfo  {journal} {J.
  Phys.: Condens. Matter}\ }\textbf {\bibinfo {volume} {22}},\ \bibinfo {pages}
  {076003} (\bibinfo {year} {2010})}\BibitemShut {NoStop}%
\bibitem [{\citenamefont {Romero}\ \emph {et~al.}(2013)\citenamefont {Romero},
  \citenamefont {Aligia}, \citenamefont {Sereni},\ and\ \citenamefont
  {Nieva}}]{Romero:2013}%
  \BibitemOpen
  \bibfield  {author} {\bibinfo {author} {\bibfnamefont {M.~A.}\ \bibnamefont
  {Romero}}, \bibinfo {author} {\bibfnamefont {A.~A.}\ \bibnamefont {Aligia}},
  \bibinfo {author} {\bibfnamefont {J.~G.}\ \bibnamefont {Sereni}}, \ and\
  \bibinfo {author} {\bibfnamefont {G.}~\bibnamefont {Nieva}},\ }\href
  {\doibase 10.1088/0953-8984/26/2/025602} {\bibfield  {journal} {\bibinfo
  {journal} {J. Phys.: Condens. Matter}\ }\textbf {\bibinfo {volume} {26}},\
  \bibinfo {pages} {025602} (\bibinfo {year} {2013})}\BibitemShut {NoStop}%
\bibitem [{\citenamefont {Kamikawa}\ \emph {et~al.}(2015)\citenamefont
  {Kamikawa}, \citenamefont {Ishii}, \citenamefont {Noguchi}, \citenamefont
  {Goto}, \citenamefont {Fujita}, \citenamefont {Nakagawa}, \citenamefont
  {Tanida}, \citenamefont {Sera},\ and\ \citenamefont
  {Suzuki}}]{Kamikawa:2015}%
  \BibitemOpen
  \bibfield  {author} {\bibinfo {author} {\bibfnamefont {S.}~\bibnamefont
  {Kamikawa}}, \bibinfo {author} {\bibfnamefont {I.}~\bibnamefont {Ishii}},
  \bibinfo {author} {\bibfnamefont {Y.}~\bibnamefont {Noguchi}}, \bibinfo
  {author} {\bibfnamefont {H.}~\bibnamefont {Goto}}, \bibinfo {author}
  {\bibfnamefont {T.~K.}\ \bibnamefont {Fujita}}, \bibinfo {author}
  {\bibfnamefont {F.}~\bibnamefont {Nakagawa}}, \bibinfo {author}
  {\bibfnamefont {H.}~\bibnamefont {Tanida}}, \bibinfo {author} {\bibfnamefont
  {M.}~\bibnamefont {Sera}}, \ and\ \bibinfo {author} {\bibfnamefont
  {T.}~\bibnamefont {Suzuki}},\ }\href {\doibase 10.1016/j.phpro.2015.12.024}
  {\bibfield  {journal} {\bibinfo  {journal} {Phys. Procedia}\ }\textbf
  {\bibinfo {volume} {75}},\ \bibinfo {pages} {187} (\bibinfo {year}
  {2015})}\BibitemShut {NoStop}%
\bibitem [{\citenamefont {M\"uller}\ \emph {et~al.}(2015)\citenamefont
  {M\"uller}, \citenamefont {Desilets-Benoit}, \citenamefont {Gauthier},
  \citenamefont {Lapointe}, \citenamefont {Bianchi}, \citenamefont {Maris},
  \citenamefont {Zahn}, \citenamefont {Beyer}, \citenamefont {Green},
  \citenamefont {Wosnitza}, \citenamefont {Yamani},\ and\ \citenamefont
  {Kenzelmann}}]{Muller:2015}%
  \BibitemOpen
  \bibfield  {author} {\bibinfo {author} {\bibfnamefont {R.~A.}\ \bibnamefont
  {M\"uller}}, \bibinfo {author} {\bibfnamefont {A.}~\bibnamefont
  {Desilets-Benoit}}, \bibinfo {author} {\bibfnamefont {N.}~\bibnamefont
  {Gauthier}}, \bibinfo {author} {\bibfnamefont {L.}~\bibnamefont {Lapointe}},
  \bibinfo {author} {\bibfnamefont {A.~D.}\ \bibnamefont {Bianchi}}, \bibinfo
  {author} {\bibfnamefont {T.}~\bibnamefont {Maris}}, \bibinfo {author}
  {\bibfnamefont {R.}~\bibnamefont {Zahn}}, \bibinfo {author} {\bibfnamefont
  {R.}~\bibnamefont {Beyer}}, \bibinfo {author} {\bibfnamefont
  {E.}~\bibnamefont {Green}}, \bibinfo {author} {\bibfnamefont
  {J.}~\bibnamefont {Wosnitza}}, \bibinfo {author} {\bibfnamefont
  {Z.}~\bibnamefont {Yamani}}, \ and\ \bibinfo {author} {\bibfnamefont
  {M.}~\bibnamefont {Kenzelmann}},\ }\href {\doibase
  10.1103/PhysRevB.92.184432} {\bibfield  {journal} {\bibinfo  {journal} {Phys.
  Rev. B}\ }\textbf {\bibinfo {volume} {92}},\ \bibinfo {pages} {184432}
  (\bibinfo {year} {2015})}\BibitemShut {NoStop}%
\bibitem [{\citenamefont {Santini}\ \emph {et~al.}(1999)\citenamefont
  {Santini}, \citenamefont {L\'{e}manski},\ and\ \citenamefont
  {Erd\H{o}s}}]{Santini:1999}%
  \BibitemOpen
  \bibfield  {author} {\bibinfo {author} {\bibfnamefont {P.}~\bibnamefont
  {Santini}}, \bibinfo {author} {\bibfnamefont {R.}~\bibnamefont
  {L\'{e}manski}}, \ and\ \bibinfo {author} {\bibfnamefont {P.}~\bibnamefont
  {Erd\H{o}s}},\ }\href {\doibase 10.1080/000187399243419} {\bibfield
  {journal} {\bibinfo  {journal} {Adv. Phys.}\ }\textbf {\bibinfo {volume}
  {48}},\ \bibinfo {pages} {537} (\bibinfo {year} {1999})}\BibitemShut
  {NoStop}%
\bibitem [{\citenamefont {Tran}\ and\ \citenamefont {\'{S}wiatek
  Tran}(2008)}]{Tran:2008}%
  \BibitemOpen
  \bibfield  {author} {\bibinfo {author} {\bibfnamefont {V.~H.}\ \bibnamefont
  {Tran}}\ and\ \bibinfo {author} {\bibfnamefont {B.}~\bibnamefont {\'{S}wiatek
  Tran}},\ }\href {\doibase 10.1039/B806728G} {\bibfield  {journal} {\bibinfo
  {journal} {Dalton Trans.}\ ,\ \bibinfo {pages} {4860}} (\bibinfo {year}
  {2008})}\BibitemShut {NoStop}%
\bibitem [{\citenamefont {Rams}\ \emph {et~al.}(2017)\citenamefont {Rams},
  \citenamefont {Tomkowicz}, \citenamefont {B\"{o}hme}, \citenamefont {Plass},
  \citenamefont {Suckert}, \citenamefont {Werner}, \citenamefont {Jess},\ and\
  \citenamefont {N\"{a}ther}}]{Rams:2017}%
  \BibitemOpen
  \bibfield  {author} {\bibinfo {author} {\bibfnamefont {M.}~\bibnamefont
  {Rams}}, \bibinfo {author} {\bibfnamefont {Z.}~\bibnamefont {Tomkowicz}},
  \bibinfo {author} {\bibfnamefont {M.}~\bibnamefont {B\"{o}hme}}, \bibinfo
  {author} {\bibfnamefont {W.}~\bibnamefont {Plass}}, \bibinfo {author}
  {\bibfnamefont {S.}~\bibnamefont {Suckert}}, \bibinfo {author} {\bibfnamefont
  {J.}~\bibnamefont {Werner}}, \bibinfo {author} {\bibfnamefont
  {I.}~\bibnamefont {Jess}}, \ and\ \bibinfo {author} {\bibfnamefont
  {C.}~\bibnamefont {N\"{a}ther}},\ }\href {\doibase 10.1039/C6CP08193B}
  {\bibfield  {journal} {\bibinfo  {journal} {Phys. Chem. Chem. Phys.}\
  }\textbf {\bibinfo {volume} {19}},\ \bibinfo {pages} {3232} (\bibinfo {year}
  {2017})}\BibitemShut {NoStop}%
\bibitem [{\citenamefont {Fu}\ \emph {et~al.}(2010)\citenamefont {Fu},
  \citenamefont {K\"{o}gerler}, \citenamefont {R\"{u}cker}, \citenamefont {Su},
  \citenamefont {Mittal},\ and\ \citenamefont {Br\"{u}ckel}}]{Fu:2010}%
  \BibitemOpen
  \bibfield  {author} {\bibinfo {author} {\bibfnamefont {Z.-D.}\ \bibnamefont
  {Fu}}, \bibinfo {author} {\bibfnamefont {P.}~\bibnamefont {K\"{o}gerler}},
  \bibinfo {author} {\bibfnamefont {U.}~\bibnamefont {R\"{u}cker}}, \bibinfo
  {author} {\bibfnamefont {Y.}~\bibnamefont {Su}}, \bibinfo {author}
  {\bibfnamefont {R.}~\bibnamefont {Mittal}}, \ and\ \bibinfo {author}
  {\bibfnamefont {T.}~\bibnamefont {Br\"{u}ckel}},\ }\href {\doibase
  10.1088/1367-2630/12/8/083044} {\bibfield  {journal} {\bibinfo  {journal}
  {New J. Phys.}\ }\textbf {\bibinfo {volume} {12}},\ \bibinfo {pages} {083044}
  (\bibinfo {year} {2010})}\BibitemShut {NoStop}%
\bibitem [{\citenamefont {Ahmed}\ \emph {et~al.}(2015)\citenamefont {Ahmed},
  \citenamefont {Tsirlin},\ and\ \citenamefont {Nath}}]{Ahmed:2015}%
  \BibitemOpen
  \bibfield  {author} {\bibinfo {author} {\bibfnamefont {N.}~\bibnamefont
  {Ahmed}}, \bibinfo {author} {\bibfnamefont {A.~A.}\ \bibnamefont {Tsirlin}},
  \ and\ \bibinfo {author} {\bibfnamefont {R.}~\bibnamefont {Nath}},\ }\href
  {\doibase 10.1103/PhysRevB.91.214413} {\bibfield  {journal} {\bibinfo
  {journal} {Phys. Rev. B}\ }\textbf {\bibinfo {volume} {91}},\ \bibinfo
  {pages} {214413} (\bibinfo {year} {2015})}\BibitemShut {NoStop}%
\end{thebibliography}
\end{document}